\documentclass[12pt]{iopart}
\usepackage{graphicx}
\usepackage{bm}
\usepackage{url}
\usepackage[active]{srcltx}

\begin{document}

%opening
%\title{Adiabatic electron-hole junctions in graphene}
\title{Chiral tunneling in single and bilayer graphene}
\author{T.~Tudorovskiy, K.~J.~A.~Reijnders, M.~I.~Katsnelson}
\address{Radboud University Nijmegen, Institute for Molecules and Materials, Heyendaalseweg 135, 6525 AJ Nijmegen, The Netherlands}
\ead{m.katsnelson@science.ru.nl}

\maketitle

\begin{abstract}
We review chiral (Klein) tunneling in single-layer and bilayer graphene and 
present its semiclassical theory, including the Berry phase
and the Maslov index. Peculiarities of the chiral tunneling are naturally
explained in terms of classical phase space. In a one-dimensional geometry we reduced
the original Dirac equation, describing the dynamics of charge carriers in the single layer graphene,
to an effective Schr\"odinger equation with a complex potential. This allowed us to study tunneling in details
and obtain analytic formulas. Our predictions are compared with numerical results.
We have also demonstrated that, for the case of asymmetric n-p-n junction in single layer graphene,
there is total transmission for normal incidence only, side resonances are suppressed.
\end{abstract}

\section{Introduction}

Since this paper is prepared for the proceedings of the Nobel symposium on graphene
we do not start with general explanations what graphene is and
why it is important, it will be very well described in other presentations.
We just refer to reviews \cite{r1,r2,r3,r4,r5,r6,r7}. Our particular subject
is chiral, or Klein (as it was called in \cite{KatsnelsonNovGeim2006}) tunneling.
This is one of the key phenomena determining peculiar electronic properties of graphene.
In light of possible applications, the Klein tunneling protects high charge carrier
mobility despite unavoidable inhomogeneities. At the same time, due to the Klein
tunneling graphene electronics cannot copy the standard semiconductor one: if you
make graphene transistor based on n-p-n junction just like for silicon, it will not be
efficient since you will not be able to lock it. These two remarks illustrate the importance
of the subject, the more detailed discussion is presented below.

The paper consists of two pieces. The first one (Sections \ref{KleinParadox}~--~\ref{KleinBi}) preserves the historical
line of thoughts and presents the motivation of the problem, from \cite{Klein1929} to \cite{KatsnelsonNovGeim2006}.
In the second part (Sections \ref{sec::dimvar}~--~\ref{sec:numerics}) we present a systematic semiclassical theory of the
chiral tunneling, together with numerical results.

  \section{\label{KleinParadox}The Klein paradox}
  Soon after the discovery of the Dirac equation, O. Klein~\cite{Klein1929} noticed one of its strange properties which was afterwards called the ``Klein paradox''. Klein considered the original four by four Dirac equation, which governs the dynamics of a spin one half particle moving in three-dimensional space. To make a direct connection to the case of graphene without changing the essence of the paradox, we will consider a two by two matrix equation for a particle propagating in two-dimensional space:
  \begin{equation}
    \hat H \Psi = E \Psi \; , \label{eq:statSchrod}
  \end{equation}
  where $\Psi = (\psi_1,\psi_2)$ and the Hamiltonian
  \begin{equation}
    \hat H = c \, \boldsymbol\sigma \hat{\boldsymbol p} + u(x,y) + m c^2 \hat{\sigma}_z \; .
  \end{equation}
  Here $m$ is the mass of the particle, $c$ is the speed of light and $u(x,y)$ is the potential energy.

  To demonstrate the essence of the paradox we consider normal incidence on a one-dimensional potential barrier, which means that $u=u(x)$ and $\psi_i=\psi_i(x)$. Then equation~(\ref{eq:statSchrod}) takes the form
  \begin{equation}
    \left\{
    \begin{array}{l}
      -i \hbar c \displaystyle{\frac{d\psi_2}{d x}} = \left(E - m c^2 - u(x) \right) \psi_1 \; , \\
      -i \hbar c \displaystyle{\frac{d\psi_1}{d x}} = \left(E + m c^2 - u(x) \right) \psi_2 \; .
    \end{array}
    \right. \label{eq:TwoDDirac}
  \end{equation}
  To make the problem exactly solvable we use a step-wise potential
  \begin{equation}
    u(x) =
    \left\{
    \begin{array}{ll}
      0, & x < 0, \\
      u_0, & x > 0,
    \end{array}\right.
  \end{equation}
  where $u_0$ is a positive constant. We consider a general scattering problem with an incoming wave $\Psi_{in}(x)$ and a reflected wave $\Psi_r(x)$ for $x<0$,
  \begin{equation}
    \Psi(x) = \Psi_{in}(x) + r \Psi_r(x) \; ,
  \end{equation}
  and a transmitted wave $\Psi_t(x)$ for $x>0$,
  \begin{equation}
    \Psi(x) = t \Psi_t(x) \; .
  \end{equation}

  The $x$-dependence of the solutions for $x<0$ can be written as $\exp(\pm i k x)$, where the wave vector $k$ satisfies the relativistic dispersion relation $E^2 = \hbar^2 c^2 k^2 + m^2 c^4$ as can be found by diagonalizing equation~(\ref{eq:TwoDDirac}) with $u=0$. Alternatively the wave vector can be written as
  \begin{equation}
    k = \frac{\sqrt{E^2 - m^2 c^4}}{\hbar c} \; .
  \end{equation}
  One easily sees that there are three distinct regimes, two of which are classically allowed, namely $E > m c^2$ corresponding to electron states and $E < - m c^2$ corresponding to hole or positron states. There is also a classically forbidden region $-m c^2 < E < m c^2$ where the wave vector $k$ is imaginary and we have evanescent waves. In what follows we will assume that we are in the electron regime. By calculating eigenvectors of equation~(\ref{eq:TwoDDirac}) one obtains for the wavefunctions to the left of the barrier
  \begin{equation}
    \Psi_{in}(x) = \left( \begin{array}{c} 1 \\ \alpha \end{array} \right) e^{i k x}
  \end{equation}
  and
  \begin{equation}
    \Psi_{r}(x) = \left( \begin{array}{c} 1 \\ -\alpha \end{array} \right) e^{-i k x} \; ,
  \end{equation}
  where
  \begin{equation}
    \alpha = \sqrt{\frac{E-m c^2}{E+m c^2}} \; . \label{eq:defalpha}
  \end{equation}

  To the right of the barrier we have a new wave vector $q$, which satisfies the relativistic dispersion relation $(E-u_0)^2 = \hbar^2 c^2 q^2 + m^2 c^4 $, or
  \begin{equation}
    q = \frac{\sqrt{(u_0-E)^2 - m^2 c^4}}{\hbar c} \; .
  \end{equation}
  Consider a jump 
  \begin{equation}
    u_0 > E + m c^2 \; , \label{eq:Kleinhighpot}
  \end{equation}
  since in this case the paradox arises. The wave vector $q$ is real and we have a propagating wave on the right side of the barrier. Note however that this particle belongs to the hole continuum rather than to the electron one. For smaller values of $u_0$, there are either propagating electrons on both the left and the right side of the barrier, when $u_0 < E - m c^2$, or evanescent waves on the right side of the barrier, when $E - m c^2 < u_0 < E + m c^2$. Solving the Dirac equation~(\ref{eq:TwoDDirac}) on the right side of the barrier, one obtains for the transmitted wave
  \begin{equation}
    \Psi_t(x) = \left( \begin{array}{c} 1 \\ -1/\beta \end{array} \right) e^{i q x} \; , \label{eq:Kleintransmitted}
  \end{equation}
  where
  \begin{equation}
    \beta = \sqrt{\frac{u_0 - E - m c^2}{u_0 - E + m c^2}}. \label{eq:defbeta}
  \end{equation}

  From the continuity of the wavefunction at $x=0$,
  \begin{equation}
    \Psi_{in} + r \Psi_r |_{x=-0} = \Psi_t |_{x=+0} \; ,
    \label{eq:continuity}
  \end{equation}
  we find
  \begin{equation}
    r = \frac{\alpha \beta + 1}{\alpha \beta - 1} \; .
  \end{equation}
  For the considered case we have $0 < \alpha, \beta < 1$, so that $r < 0$ and
  \begin{equation}
    R = |r|^2 = \left( \frac{1 + \alpha \beta}{1 - \alpha \beta} \right)^2 > 1 \; .
  \end{equation}
   To treat reflection and transmission coefficients properly one has to look at the probability current density for the one-dimensional Dirac equation
  \begin{equation}
    j_x = c \Psi^{\dagger} \sigma_x \Psi = c(\psi_1^* \psi_2 + \psi_2^* \psi_1) \; , \label{eq:currdensity}
  \end{equation}
  which is a conserved quantity.   When we look at the current density~(\ref{eq:currdensity})   
   we see that it takes values $2 \alpha c$ for the incoming wave and $-2 \alpha c R$ for the reflected wave. Therefore $R$ is nothing but the reflection coefficient and we come to the conclusion that the amplitude of the reflected wave is larger than the amplitude of the incident one. This strange effect that occurs when condition~(\ref{eq:Kleinhighpot}) is fulfilled was initially called the Klein paradox. In our further discussion we will follow~\cite{CaloDomb1999} and~\cite{DombCalo1999}. For a rather complete list of references see~\cite{GreinerSchram2008}.

  First of all note that the current density~(\ref{eq:currdensity}) on the right hand side equals $-2 |t|^2/\beta$, indicating that there is something wrong with the definition of the transmitted wave. What exactly is wrong was pointed out by Pauli, who noticed that the group velocity for the case of equation~(\ref{eq:Kleinhighpot}),
  \begin{equation}
    v_g = \frac{1}{\hbar} \frac{d E}{d q} = \frac{1}{\hbar} \left(\frac{d q}{d E} \right)^{-1} = \frac{\hbar c^2 q}{E - u_0} \; ,
    \label{eq:groupvelocity}
  \end{equation}
  is opposite to the direction of the wave vector $q$. Since the group velocity determines the direction of propagation, the transmitted wave~(\ref{eq:Kleintransmitted}) corresponds (for positive $q$) to a particle moving to the left instead of to the right. Therefore we should define our outgoing wave as
  \begin{equation}
    \Psi_t(x) = \left( \begin{array}{c} 1 \\ 1/\beta \end{array} \right) e^{-i q x} \; ,
  \end{equation}
  which gives the currenty density $2 |t|^2/\beta$. When we once again calculate $r$ from equation~(\ref{eq:continuity}), it is seen that
  \begin{equation}
    R = |r|^2 = \left( \frac{1 - \alpha \beta}{1 + \alpha \beta} \right)^2 < 1 \; .
    \label{eq:Kleinreflected}
  \end{equation}
  which is always smaller than one. Therefore the formal paradox disappears, see also~\cite{VonsovskySvirski1993}.

  The paradox reappears when we consider the problem from a different angle. Instead  of an infinitely broad barrier we will consider a finite barrier,
  \begin{equation}
    u(x) =
    \left\{
    \begin{array}{ll}
      u_0, & |x| < a \\
      0, & |x| > a
    \end{array}\right.
    \label{eq:potentialbarrier}
  \end{equation}
  The problem with the choice of the transmitted wave on the right side of the barrier has now disappeared, since it is simply $t \Psi_{in}$. Within the barrier one now has to consider both modes $\exp(\pm i q x)$, representing the most general solution. Reflection and transmission coefficients are then obtained from the continuity of the wave function at $x=-a$ and $x=a$, which gives after straightforward calculations (see e.g.~\cite{CaloDomb1999} and~\cite{SuSiuChou1993})
  \begin{eqnarray}
    R &=& \frac{(1-\alpha^2\beta^2)^2 \sin^2(2 q a)}{4 \alpha^2 \beta^2 + (1-\alpha^2 \beta^2)^2 \sin^2(2 q a)}, \\
    T &=& \frac{4 \alpha^2 \beta^2}{4 \alpha^2 \beta^2 + (1-\alpha^2 \beta^2)^2 \sin^2(2 q a)}.
  \end{eqnarray}
  There is no paradox in these expressions, since $0 < R < 1$, $0 < T < 1$ and $R+T=1$ as it should be. Note that we have total transmission through the barrier when
  \begin{equation}
    q a = \frac{N \pi}{2} \; ,
  \end{equation}
  with integer $N$.

  We can consider an infinitely broad barrier by letting $a$ go to infinity in the above expressions.  As $a$ becomes very large while other parameters remain fixed the sine will oscillate very rapidly. We can then average over the fast oscillations and replace $\sin^2(2 q a)$ by its average value $\frac{1}{2}$ to obtain the expressions
  \begin{eqnarray}
    R_{\infty} &=& \frac{(1-\alpha^2\beta^2)^2}{8 \alpha^2 \beta^2 + (1-\alpha^2 \beta^2)^2} \label{eq:Kleinreflected2} \\
    T_{\infty} &=& \frac{8 \alpha^2 \beta^2}{8 \alpha^2 \beta^2 + (1-\alpha^2 \beta^2)^2} \label{eq:Kleintransmitted2}
  \end{eqnarray}
  One may be surprised that the results~(\ref{eq:Kleinreflected}) and~(\ref{eq:Kleinreflected2}) do not coincide. It is however well known from electromagnetic wave theory~\cite{Stratton} that the reflection coefficients for the two situations should differ.

  From the last result we see once again that the paradox has disappeared in its mathematical form, but has reappeared as physically counterintuitive behaviour. In non-relativistic quantum mechanics a particle can tunnel through a classically forbidden region $E < u(x)$, but the probability is exponentially small when the barrier is high and broad. In the semiclassical approximation the transmission through the barrier with turning points $x_{1,2}$, which satisfy $E=u(x_{1,2})$, is given by
  \begin{equation}
    T = \exp\left(-\frac{2}{\hbar}\int_{x_1}^{x_2} dx \sqrt{2 m (u(x) - E)}\right) \; ,
  \end{equation}
  where $m$ is the mass of the particle. For a relativistic particle incident on a sufficiently high barrier~(\ref{eq:Kleinhighpot}) the situation is dramatically different. In the limit $a \to \infty$ the probability of penetration~(\ref{eq:Kleintransmitted2}) is in general not small at all. Even for an infinitely high barrier ($u_0 \to \infty$) one has $\beta = 1$ and
  \begin{equation}
    T_{\infty} = \frac{E^2 - m^2 c^4}{E^2 - \frac{1}{2} m^2 c^4} \; .
  \end{equation}
  This is of the order of one when $E- m c^2$ is of the order of $m c^2$, while it is approximately equal to one in the ultrarelativistic limit
  \begin{equation}
    E \gg m c^2 \; .
  \end{equation}
  This is the contemporary formulation of the Klein paradox~\cite{CaloDomb1999}; quantum relativistic particles can tunnel with large enough probabilities through barriers of arbitrarily large height and width.

  The tunneling effect can be hand-wavingly explained with the help of the Heisenberg uncertainty principle. Since one cannot know both momentum and position with an arbitrary accuracy at a given instant, one cannot separate the total energy into a potential and a kinetic part. So the kinetic energy can be ``a bit'' negative. In the relativistic regime the restriction is much stronger~\cite{LandauPeierls1931}: one cannot even know the coordinate with an accuracy higher than $\hbar c /E$. Therefore relativistic quantum mechanics cannot be mechanics, but can only be field theory~\cite{Lifshitz1971}. This theory will always contain particles and antiparticles and to measure the coordinate better than $\hbar c / E$ one needs to apply such a high energy that particle-antiparticle pairs will be created. The original particle whose coordinate one wanted to measure will then be lost among the newly-born particles. A full field theoretic treatment of the problem was given in Ref.~\cite{CaloDombImagawa1995}. The most important point is that although the problem of a high enough barrier looks like a static problem, this is actually not the case. One needs to study carefully how the state is reached and this involves positron emission by the growing barrier. For a more detailed discussion of the role of electron-positron pairs in the Klein paradox, see~\cite{KrekSuGrobe2005}.

  \section{\label{sec::KleinSingle}Klein tunneling in single layer graphene}
  The Hamiltonian for charge carriers in graphene near conical points $K$ and $K'$ is given by the massless Dirac Hamiltonian
  \begin{equation}
    \hat H = V \left(  \sigma_x \hat{p}_x + \sigma_y \hat{p}_y \right) + u(x,y) \; ,
    \label{eq:grapheneHamiltonian}
  \end{equation}
  where $V$ is the Fermi velocity $V \approx c/300$. To consider normal incidence on the one-dimensional potential barrier~(\ref{eq:potentialbarrier}) in this case, we can simply put $m=0$ in our previous results. From equations~(\ref{eq:defalpha}) and~(\ref{eq:defbeta}) it is seen that $\alpha = \beta = 1$. Therefore we have $T = 1$ and $R=0$ in equations~(\ref{eq:Kleinreflected2}) and~(\ref{eq:Kleintransmitted2}), regardless of the height of the potential. This result is not related to the specific shape of the potential~\cite{Ando1998}.

  This property has an analog in two and three dimensions with $u=u(x,y)$ or $u=u(x,y,z)$, namely that backscattering is forbidden. This was found long ago for scattering of ultrarelativistic particles in three dimensions (see ~\cite{Yennie1954, Lifshitz1971}). An important consequence of this property for carbon materials was noticed in~\cite{Ando1998}. Absence of backscattering explains the existence of conducting channels in metallic carbon nanotubes, while in a non-relativistic one-dimensional system an arbitrarily small disorder leads to localization~\cite{Lifshitz1988}.

  The consideration in~\cite{Ando1998} is very instructive since it explicitly shows the role of the Berry phase and time-reversal symmetry, but it is also quite cumbersome. Here we present a somewhat simplified scheme of this proof. To this aim we consider the equation for the $T$-matrix (see e.g.~\cite{Newton1966})
  \begin{equation}
    \hat{T} = \hat{u} + \hat{u}\hat{G}_0\hat{T} \; ,
    \label{eq:defTmatrix}
  \end{equation}
  where $\hat{u}$ is the operator corresponding to the scattering potential,
  \begin{equation}
    \hat{G}_0 = \lim_{\delta \to +0} \frac{1}{E - \hat{H}_0 + i \delta} \; ,
  \end{equation}
  is the Green's function of the unperturbed Hamiltonian $\hat{H}_0$ and $E$ is the electron energy, which is assumed to be larger than zero. If $\hat{H}_0$ is the Dirac Hamiltonian for massless Dirac fermions~(\ref{eq:grapheneHamiltonian}), we have
  \begin{equation}
    \hat{G}_0(\mathbf{r},\mathbf{r'}) = \int \frac{d\mathbf{q}}{(2\pi)^2} \hat{G}_0(\mathbf{q}) \exp[i \mathbf{q} (\mathbf{r}-\mathbf{r}')]\; ,
  \end{equation}
  where
  \begin{equation}
    \hat{G}_0(\mathbf{q}) = \frac{1}{E-\hbar V \mathbf{q}\boldsymbol\sigma + i \delta} =
    \frac{1}{\hbar V} \frac{\varepsilon + \mathbf{q}\boldsymbol\sigma}{(\varepsilon+i\delta)^2-q^2} \; ,
  \end{equation}
  with $\varepsilon = E/\hbar V$. The probability of backscattering can be found by iterating equation~(\ref{eq:defTmatrix}) and is proportional to
  \begin{equation}
    T(-\mathbf{k},\mathbf{k}) = \left\langle -\mathbf{k} \left| u + u\hat{G}_0 u + u\hat{G}_0 u \hat{G}_0 u + \ldots \right| \mathbf{k} \right\rangle \equiv T^{(1)} + T^{(2)} + \ldots \; , 
    \label{eq:Tmatrixexpansion}
  \end{equation}
  where $T^{(n)}$ is the contribution proportional to $u^n$.

  We can always choose axes such that $\mathbf{k} \parallel Ox$. In this case $\left| \mathbf{k} \right\rangle$ and $\left| -\mathbf{k} \right\rangle$ have spinor structures $\left( \begin{array}{c} 1 \\ 1 \end{array} \right)$ and $\left( \begin{array}{c} 1 \\ -1 \end{array} \right)$ respectively. Therefore, if $\hat{T}$ is the two by two matrix
  \begin{equation}
    \hat{T} = T_0 + \boldsymbol T \boldsymbol\sigma \; ,
  \end{equation}
  one has
  \begin{equation}
    T(-\mathbf{k},\mathbf{k}) \sim T_z(-\mathbf{k},\mathbf{k}) + i T_y(-\mathbf{k},\mathbf{k})
  \end{equation}
  Now keeping in mind that $V$ is proportional to the identity matrix one can prove term by term that all contributions to $T_y(-\mathbf{k},\mathbf{k})$ and $T_z(-\mathbf{k},\mathbf{k})$ vanish by symmetry. Actually this is because $\hat{\mathbf{T}}(\mathbf{k}) \sim \mathbf{k} \parallel Ox$; from the vectors $\mathbf{k}$ and $-\mathbf{k}$ one cannot construct anything with nonzero $y$ or $z$ components. Strictly speaking this argument is only enough for an isotropic potential; for a generic case one has to do a term by term analysis based on expansion~(\ref{eq:Tmatrixexpansion}), see Ref.~\cite{Ando1998}. For two nonparallel vectors $\mathbf{k_1}$ and $\mathbf{k_2}$ one can construct a matrix with nonzero $y$ or $z$ components, since one of the vectors has a nonzero $y$ component, so that $\mathbf{k_1} \times \mathbf{k_2} \parallel Oz$.

  When one thinks about electrons in quantum electrodynamics, it is not easy to create potential jumps larger than $2 m c^2 \approx 1$ MeV. Similar phenomena take place in electric or gravitational fields (\cite{GreinerMuellerRafaelski1985, GribMamaevMostepanenko1994}; see~\cite{GreinerSchram2008} for a detailed list of references), but the context is always quite exotic, such as collisions of ultraheavy ions or even black hole evaporation. There were no experimental data available which would require the Klein paradox for their explanation. However shortly after the discovery of graphene it was realized that Klein tunneling is one of the crucial phenomena for graphene physics and electronics~\cite{KatsnelsonNovGeim2006}. Soon after this theoretial prediction the effect was confirmed experimentally \cite{StanderHuardGoldhaberGordon2009, YoungKim2009}.

  Considering possible applications, Klein tunneling in graphene is rather bad news. If one copied the construction from a silicon transistor to graphene, it would be impossible to lock the transistor. One would need to open a gap in the spectrum to be able to lock it. At the same time it is good news as well: due to the Klein paradox inhomogeneities in the electron density do not lead to localization and their effect on the electron mobility is not very essential~\cite{KatsnelsonNovGeim2006}.

  \section{Tunneling trough a stepwise barrier} \label{sec:singlelayerstep}
  Let us now consider a massless Dirac fermion incident on the potential barrier~(\ref{eq:potentialbarrier}) with positive energy under an angle $\phi$, as it was done first in~\cite{KatsnelsonNovGeim2006}. Of course, the potential cannot be sharp on the atomic scale, since this would induce Umklapp scattering between different valleys. Therefore by a step-wise potential we mean that the electron wavelength $k^{-1}$ is much larger than the typical spatial scale of the potential $l$, which is in turn much larger than the size of the unit cell.
  
  Within this assumption the solution is each region is given by traveling waves proportional to $\exp(\pm i k_x x) \exp(\pm i k_y y)$, where $k_x$ and $k_y$ satisfy the dispersion relation
  \begin{equation}
    \left( \frac{E-u_0}{\hbar V} \right)^2 \equiv k^2 = (k_x^2 + k_y^2) \; ,
  \end{equation}
  as can be found from equation~(\ref{eq:grapheneHamiltonian}). Similarly to the original Dirac equation we can distinguish three distinct regimes from this equation. For $ u_0 < E - \hbar V |k_y|$ we have electrons and for $u_0 > E + \hbar V |k_y|$ we have holes, while the region $E - \hbar V |k_y| < u_0 < E + \hbar V |k_y|$ is classically forbidden. As was done for the case of the massive Dirac equation, we will now require that the potential $u_0$ in equation~(\ref{eq:potentialbarrier}) satisfies
  \begin{equation}
    u_0 > E + \hbar V |k_y| \; ,
  \end{equation}
  so that we have hole states within the barrier.

  Let us denote by $k$ the wave vector for $|x| > a$ and by $q$ the wave vector for $|x| < a$. At the potential jump the momentum in the $y$ direction should be conserved, so that the new angle $\theta$ is related to the new wave vector $q$ by
  \begin{equation}
    k \sin \phi = k_y = q_y = q \sin\theta \; .
  \end{equation}
  From equation~(\ref{eq:grapheneHamiltonian}) we see that the second component of the wavefunction is related to the first by
  \begin{equation}
    \psi_2 = \mathrm{sgn}(E-u_0) e^{i \phi} \psi_1 \;,
  \end{equation}
  so the solutions in the three regions are given by
  \begin{equation}
 \fl
    \Psi(x,y) =
      \left\{
      \begin{array}{ll}
      \left( \begin{array}{c} 1 \\ s e^{i \phi} \end{array} \right) e^{i k_x x} e^{i k_y y} +
      r \left( \begin{array}{c} 1 \\ - s e^{-i \phi} \end{array} \right) e^{-i k_x x} e^{i k_y y} , & x < -a \\
      A \left( \begin{array}{c} 1 \\ s' e^{i \theta} \end{array} \right) e^{i q_x x} e^{i k_y y} +
      B \left( \begin{array}{c} 1 \\ - s' e^{-i \theta} \end{array} \right) e^{-i q_x x} e^{i k_y y} , & -a < x < a \\
      t \left( \begin{array}{c} 1 \\ s e^{i \phi} \end{array} \right) e^{i k_x x} e^{i k_y y}, & x > a
    \end{array}\right. \label{eq:singlelayerconstwave}
  \end{equation}
  where we have introduced $s = \mathrm{sgn}(E)$, $s' = \mathrm{sgn}(E-u_0)$, $k_x = k \cos\phi$ and $q_x = q \cos\theta$. Note that the reflected particle moves under the angle $\pi - \phi$, assuming that the angle changes from $-\pi/2$ to $3\pi/2$, so that we have the phase $-\exp(-i \phi)$ for the reflected wave. We can now determine the reflection coefficient $r$, the transmission coefficient $t$ and the coefficients $A$ and $B$ as before, from the requirement that the wavefunction is continuous at $x=\pm a$.

  Finally the result is given by
  \begin{equation}
 \fl
    r = 2 e^{i \phi - 2 i k_x a} \sin(2 q_x a) \frac{\sin\phi - s s' \sin \theta}
    {s s' \left[e^{-2 i q_x a} \cos(\phi + \theta) + e^{2 i q_x a} \cos(\phi - \theta) \right] - 2 i \sin(2 q_x a)} \; . \label{eq:graphenereflectionbarrier}
  \end{equation}
  For the case under consideration we have $s s' = -1$, since the signs of $E$ and $E-u_0$ are opposite. The transmission probability can now easily be calculated as
  \begin{equation}
    T = |t|^2 = 1 - |r|^2 \; .
  \end{equation}
  From equation~(\ref{eq:graphenereflectionbarrier}) we immediately see that the reflection is zero for normal incidence, as we proved for a more general potential in the previous section. There are also additional angles, called ``magic angles'', at which the reflection coefficient is zero and we have full transmission. They are given by the condition
  \begin{equation}
    q_x a = N \frac{\pi}{2} \; ,
  \end{equation}
  where $N$ is an integer.

  We can compare the behaviour of electrons in single layer graphene with the behaviour of normal electrons. When the potential barrier contains no electronic states, the transmission decays exponentially with increasing barrier width and height,~\cite{Esaki1958}, so that the barrier would reflect electrons completely. But since single layer graphene is gapless, it seems more appropriate to compare it to a gapless semiconductor with non-chiral charge carriers, a situation which can be realized in certain heterostructures~\cite{Meyeretal1995, Teissieretal1996}. For this case we find
  \begin{equation}
    t = \frac{4 k_x q_x \exp(2 i q_x a)}{(q+k_x)^2 \exp(-2 i q_x a) - (q_x - k_x)^2 \exp(2 i q_x a)} \; ,
    \label{eq:normalelectrontransmission}
  \end{equation}
  where $k_x$ and $q_x$ are the $x$-components of the wave vector outside and inside the barrier, respectively. As in the case of single layer graphene there are resonance conditions at which the barrier is transparent, given by $2 q_x a = N \pi$, where $N$ is an integer. For normal incidence we see that the transmission coefficient is an oscillating function of the tunneling parameters and can exhibit any value between zero and one. This is in contrast to single layer graphene, where the transmission is always perfect.

  \section{\label{KleinBi}Klein tunneling in bilayer graphene}
  Bilayer graphene consists of two layers of graphene on top of each other, the second layer being rotated by 120 degrees with respect to the first one. In this configuration the sublattices $A$ lie exactly on top of each other and the hopping parameter $\gamma_1$ between them is approximately 0.4 eV~\cite{Dresselhaus2002,Kuzmenko09}, while the in-plane hopping parameter $\gamma_0=t$ is approximately an order of magnitude larger. When we consider only low energy excitations, $|E|, |E-u_0| \ll 2 |\gamma_1|$, the effective Hamiltonian is given by~\cite{McCannFalko2006, Novoselov2006}
  \begin{equation}
    \hat H = \left(
    \begin{array}{cc}
      0 & (\hat{p}_x - i\hat{p}_y)^2/(2m) \\
      (\hat{p}_x + i\hat{p}_y)^2/(2m) & 0
    \end{array}
    \right) + u(x)\; ,
    \label{eq:bilayergrapheneHamiltonian}
  \end{equation}
  where the effective mass $m = \gamma_1/2V^2 \approx 0.054 m_e$, $m_e$ being the free electron mass~\cite{McCannAbergelFalko2007}. There is also hopping between the $B$ sublattices of both layers, which is denoted by $\gamma_3 \approx 0.3$ eV. When we include this parameter into the description an extra term is added to the Hamiltonian, which corresponds to so-called trigonal warping. This effect is however only important for small wave vectors~\cite{McCannAbergelFalko2007}, we will exclude it assuming that $ka, qa \gg \gamma_3 \gamma_1 / \gamma_0^2$.

   Let us consider an electron incident on the potential step~(\ref{eq:potentialbarrier}) under an angle $\phi$, as was done in~\cite{KatsnelsonNovGeim2006}. Since the potential is constant in the $y$-direction we can write the solution as
  \begin{equation}
    \Psi(x,y) = \Psi(x) e^{i k_y y} \; .
  \end{equation}
  Inserting this into equation~(\ref{eq:statSchrod}) with the Hamiltonian~(\ref{eq:bilayergrapheneHamiltonian}), we obtain
  \begin{equation}
    \left( \frac{d^2}{d x^2} - k_y^2 \right)^2 \psi_i = \left( \frac{2 m (E-u)}{\hbar^2} \right)^2 \psi_i \equiv k^4 \psi_i \; .
    \label{eq:bilayercstreduced}
  \end{equation}
  The solutions are therefore given by propagating waves $\exp(\pm i k_x x)$ and exponentially growing and decaying modes $\exp(\pm\kappa_x x)$,
  \begin{eqnarray}
    k_x^2 + k_y^2 &=& \frac{2 m |E-u|}{\hbar^2} \; , \\
    \kappa_x^2 - k_y^2 &=& \frac{2 m |E-u|}{\hbar^2} \; .
  \end{eqnarray}
  The presence of evanescent modes is markedly different from both the Schr\"odinger case and the Dirac case.   Once again there are three regimes. There are electron states for $u_0 < E - \hbar^2 k_y^2/(2 m)$ and hole states for $u_0 > E + \hbar^2 k_y^2/(2 m)$, while the region in between is classically forbidden. In what follows we assume that $u_0$ in equation~(\ref{eq:potentialbarrier}) satisfies
  \begin{equation}
    u_0 > E + \frac{\hbar^2 k_y^2}{2 m} \; .
  \end{equation}
  To find the spinors that are the solutions to equation~(\ref{eq:bilayercstreduced}) we note that the components are related by
  \begin{equation}
    \left(\frac{d}{dx} + k_y \right)^2 \psi_2 = \frac{2 m (E-u)}{\hbar^2} \psi_1 \; ,
  \end{equation}
  as can be seen from the Hamiltonian~(\ref{eq:bilayergrapheneHamiltonian}).

  Now let $k = \sqrt{2 m E}/\hbar$ be the wave vector for the propagating modes in the region $|x| > a$, while $q = \sqrt{2 m (u_0 -E)}/\hbar$ is the wave vector in the region $|x| < a$. Then the solution for $x < -a$ is given by
  \begin{equation}
    \Psi(x) =
      a_1 \left( \begin{array}{c} 1 \\ s e^{2 i \phi} \end{array} \right) e^{i k_x x} +
      b_1 \left( \begin{array}{c} 1 \\ s e^{- 2 i \phi} \end{array} \right) e^{-i k_x x} +
      c_1 \left( \begin{array}{c} 1 \\ - s h_1 \end{array} \right) e^{\kappa_x x} \; , \label{eq:bilayerconstincoming}
  \end{equation}
  where $k_y = k \sin\phi$, $k_x = k \cos \phi$, $s = \textrm{sgn}(E)$, $\kappa_x = \sqrt{k_x^2 + 2 k_y^2} = k\sqrt{1+\sin^2\phi}$ and finally $h_1 = (\sqrt{1+\sin^2\phi}-\sin\phi)^2$. The amplitude $a_1$ is the amplitude for the incoming wave in this expression, while $b_1$ corresponds to the reflected wave. For $x > a$ we have the general solution
  \begin{equation}
    \Psi(x) =
      a_3 \left( \begin{array}{c} 1 \\ s e^{2 i \phi} \end{array} \right) e^{i k_x x} +
      d_3 \left( \begin{array}{c} 1 \\ - s/h_1 \end{array} \right) e^{-\kappa_x x} \; , \label{eq:bilayerconstoutgoing}
  \end{equation}
  where $a_3$ is the transmission coefficient. Inside the barrier we need the most general solution with two propagating modes and two modes with real exponentials,
  \begin{eqnarray}
     \Psi(x) &=&
      a_2 \left( \begin{array}{c} 1 \\ s' e^{2 i \theta} \end{array} \right) e^{i q_x x} +
      b_2 \left( \begin{array}{c} 1 \\ s' e^{- 2 i \theta} \end{array} \right) e^{-i q_x x} \nonumber\\
      &+&c_2 \left( \begin{array}{c} 1 \\ - s' h_2 \end{array} \right) e^{\lambda_x x} +
      d_2 \left( \begin{array}{c} 1 \\ - s'/h_2 \end{array} \right) e^{-\lambda_x x}, \label{eq:bilayerconstgeneralsol}
  \end{eqnarray}
  where $q_y = q \sin\theta = k_y$ because the transverse momentum is conserved. Furthermore $q_x = q \cos\theta$, $s'=\textrm{sgn}(E-u_0)$, $\lambda_x = q \sqrt{1+\sin^2\theta}$ and $h_2 = (\sqrt{1+\sin^2\theta}-\sin\theta)^2$.

  Now the coefficients $a_i$, $b_i$, $c_i$ and $d_i$ have to be found from the continuity of $\psi_i(x)$ and the derivative $d\psi_i/dx$ at the points $x=\pm a$. When the problem is solved numerically, one sees that the transmission probability at normal incidence is exponentially small. Similar to the case of single layer graphene, there are once again ``magic angles'' in the spectrum, at which there is total transmission. The existence of magic angles in bilayer graphene has the same consequences as in single layer graphene, meaning that we cannot lock a conventional transistor made from bilayer graphene.

  For the case of normal incidence $\phi = \theta = 0$ we can also solve the problem analytically. The transmission coefficient is given by
  \begin{equation}
    t = \frac{4 i k q \exp(2 i k a)}{(q+i k)^2\exp(- 2 q a) - (q - i k)^2\exp(2 q a)} \; ,
  \end{equation}
  which is indeed exponentially small. When we let $a$ go to infinity, the transmission probability $T = |t|^2$ becomes zero at normal incidence. Furthermore for a single n-p junction with $u_0 \gg E$ the following analytical solution can be found for any $\phi$
  \begin{equation}
    T = \frac{E}{u_0} \sin^2(2 \phi) \; ,
  \end{equation}
  which also gives $T=0$ at normal incidence, in contrast to the case of single layer graphene, where normally incident electrons are always transmitted. It is also different from the case of normal electrons, where the transmission is given by equation~(\ref{eq:normalelectrontransmission}).

\section{\label{sec::dimvar}Dimensionless variables and parameters}

In sections \ref{sec::KleinSingle}, \ref{KleinBi} it was discussed that the wavefunctions $\Psi$ of charge carriers in single layer and bilayer graphene in a one-dimensional geometry obey equations
\begin{equation}
 \left[V\left(\begin{array}{cc}0 & \hat p_x-ip_y\\ \hat p_x+ip_y & 0 \end{array}\right)+u(x/l)-E\right]\Psi=0,
\label{eq::single-dim}
\end{equation}
and
\begin{equation}
 \left[\frac{1}{2m}\left(\begin{array}{cc}0 & (\hat p_x-ip_y)^2\\ (\hat p_x+ip_y)^2 & 0 \end{array}\right)+u(x/l)-E\right]\Psi=0,
\label{eq::bi-dim}
\end{equation}
respectively. Here $l$ is a characteristic scale of a potential change. In dimensionless variables (\ref{eq::single-dim}) takes the form
\begin{equation}
 \left[\left(\begin{array}{cc}0 & \tilde p_x-i \tilde p_y\\
\tilde p_x+i \tilde p_y & 0 \end{array}\right)+\tilde u(\tilde x)-\widetilde E \right]\Psi=0,
\label{eq::single-dimless}
\end{equation}
where $\tilde x=x/l$, $\tilde p_x=-ihd/d \tilde x$, $\tilde p_y=p_y/p_0$, $h=\hbar/p_0l$, $\tilde u=u/Vp_0$ and $\widetilde E=E/Vp_0$. We denote some characteristic value of $|u-E|$ as $Vp_0$.

Analogously, (\ref{eq::bi-dim}) can be rewritten as
\begin{equation}
 \left[\left(\begin{array}{cc}0 & (\tilde p_x-i\tilde p_y)^2\\ (\tilde p_x+i\tilde p_y)^2 & 0 \end{array}\right)+\tilde u(\tilde x)-\widetilde E\right]\Psi=0,
\label{eq::bi-dimless}
\end{equation}
with $\tilde x=x/l$, $\tilde p_x=-ihd/d \tilde x$, $\tilde p_y=p_y/p_0$, $h=\hbar/p_0l$, $\tilde u=2mu/p_0^2$ and $\widetilde E=2mE/p_0^2$. We denote some characteristic value of $|u-E|$ as $p_0^2/2m$.

Thus we can introduce  dimensionless Hamiltonians (we omitted tildes):
\begin{equation}
\hat H=\left(\begin{array}{cc}0 & \hat p_x-i p_y\\
\hat p_x+i p_y & 0 \end{array}\right)+ u(x)
\label{eq::single-dimless2}
\end{equation}
for a single layer and
\begin{equation}
\hat H=\left(\begin{array}{cc}0 & (\hat p_x-i p_y)^2\\ (\hat p_x+i p_y)^2 & 0 \end{array}\right)+u(x)
\label{eq::bi-dimless2}
\end{equation}
for a bilayer. In both cases there are two substantial parameters in the problem: $h$ and $p_y$.

\section{\label{sec::standard-semi}Standard semiclassical treatment}

Charge carriers in single layer graphene are described by the Hamiltonian (\ref{eq::single-dimless2}). This Hamiltonian describes simultaneously coupled electron and hole states. According to \ref{app::ad-ham}, in adiabatic approximation (\ref{eq::single-dimless2}) can be diagonalized up to any order of $h\ll 1$. The obtained scalar Hamiltonians describe electrons and holes separately. The diagonalization is based on a series of unitary transformations of the original Hamiltonian and traces back to the ideas of the Foldy-Wouthuysen transformation \cite{Foldy50} and the Peierls substitution in Blount's treatment \cite{Blount61}. We use its variant \cite{Berlyand87,Belov06}.

Effective electron and hole Hamiltonians $\hat L^+$ and $\hat L^-$ can be written as series with respect to the small parameter $h$:
\begin{equation}
 L^\pm(\hat p_x,x,h)=L_0^\pm(\hat p_x,x)+hL_1^\pm(\hat p_x,x)+h^2L_2^\pm(\hat p_x,x)+\ldots
\end{equation}
To be precise we will assume that any function of $\hat p_x$ and $x$ is defined in such a way that $\hat p_x$ acts the first. As soon as the ordering of operators has been introduced, one can work with functions of $c$-numbers $p_x$ and $x$. These functions are called ``symbols'' \cite{Maslov72,Maslov81}.

It is shown in \ref{app::ad-ham}, that leading terms $L_0^\pm(p_x,x)$ of the effective Hamiltonians $L^\pm(p_x,x,h)$ are eigenvalues of $H(p_x,x)$:
\begin{equation}
 H(p_x,x)\chi_0^\pm(p_x,x)=L_0^\pm(p_x,x)\chi_0^\pm(p_x,x),
\end{equation}
where $\chi_0^\pm(p_x,x)$ are two eigenvectors of the matrix $H(p_x,x)$. This gives
\begin{equation}
 L_0^\pm(p_x,x)=\pm |p|+u(x), \qquad \chi_0^\pm(p_x)=\frac{1}{\sqrt{2}}\left(\begin{array}{c}e^{-i\phi_p} \\ \pm 1\end{array}\right).
\label{eq::L0-single}
\end{equation}
We note that in the absence of a magnetic field $\chi_0^\pm$ does not depend on $x$. The first correction $L_1^\pm(p_x,x)$ reads
\begin{equation}
 L_1^\pm(p_x,x)=i\left(\chi_0^\pm\right)^\dag\frac{\partial \chi_0^\pm}{\partial p_x}\frac{\partial L_0^\pm}{\partial x}=
\frac{1}{2}\frac{\partial L_0^\pm}{\partial x}\frac{\partial\phi_p}{\partial p_x}=-\frac{u'(x)}{2}\frac{p_y}{p_x^2+p_y^2}.
\end{equation}
Standard semiclassical treatment (see \ref{app::sem-appr}) can be applied to scalar Schr\"odinger-like equations $\hat L^\pm\psi^\pm=E\psi^\pm$. We are looking for a solution in the form $\psi^\pm=e^{iS^\pm(x)/h}A^\pm(x,h)$, $A^\pm(x,h)=A_0^\pm(x)+hA_1^\pm(x)+\ldots$ This gives
\begin{eqnarray}
A_0^\pm(x)=\left|\frac{\partial L_0^\pm}{\partial p_x}\right|^{-1/2}\exp\left[-i\int dx \left(\frac{\partial L_0^\pm}{\partial p_x}\right)^{-1}
\left(L_1^\pm+\frac{i}{2}\frac{\partial^2 L_0^\pm}{\partial p_x\partial x}\right) \right]
\end{eqnarray}
with $p_x=dS^\pm/dx$ to be found from the Hamilton-Jacobi equation $L_0^\pm(p_x,x)=E$, where
\begin{equation}
 \left|\frac{\partial L_0^\pm}{\partial p_x}\right|=\frac{|p_x|}{|p|}=\frac{\left([E-u(x)]^2-p_y^2\right)^{1/2}}{|E-u(x)|}.
\end{equation}
Differentiating the Hamilton-Jacobi equation with respect to $x$ we find
\begin{equation}
 \frac{\partial L_0^\pm}{\partial p_x}\frac{d p_x}{dx}+\frac{\partial L_0^\pm}{\partial x}=0,
\end{equation}
whence
\begin{equation}
 \frac{\partial L_0^\pm}{\partial p_x}=-\frac{1}{p'_x}\frac{\partial L_0^\pm}{\partial x}.
\end{equation}
This gives
%\begin{eqnarray}
%A_0^\pm(x)&=&\frac{(E-u(x))^{1/2}}{\left|(E-u(x))^2-p_y^2\right|^{1/4}}e^{i\phi_p/2}.
%\label{eq::amp-single}
%\end{eqnarray}
%
\begin{eqnarray}
\psi(x)&=&\frac{|E-u(x)|^{1/2}}{\left[(E-u(x))^2-p_y^2\right]^{1/4}}e^{\pm iS^+(x)/h+i\phi^\pm_p(x)/2}.
\label{eq::amp-single-pm}
\end{eqnarray}
Though it is possible to define locally $\chi_0^\pm$ in (\ref{eq::L0-single}) as
\begin{equation}
\chi_0^\pm(p_x)=\frac{1}{\sqrt{2}}\left(\begin{array}{c}e^{-i\phi_p/2} \\ \pm e^{i\phi_p/2}\end{array}\right)
\end{equation}
to obtain $L_1^\pm=0$, such a choice does not provide a single-valued function in the classical phase space.

\begin{figure}
\begin{center}
\parbox{5cm}{\includegraphics[height=4cm]{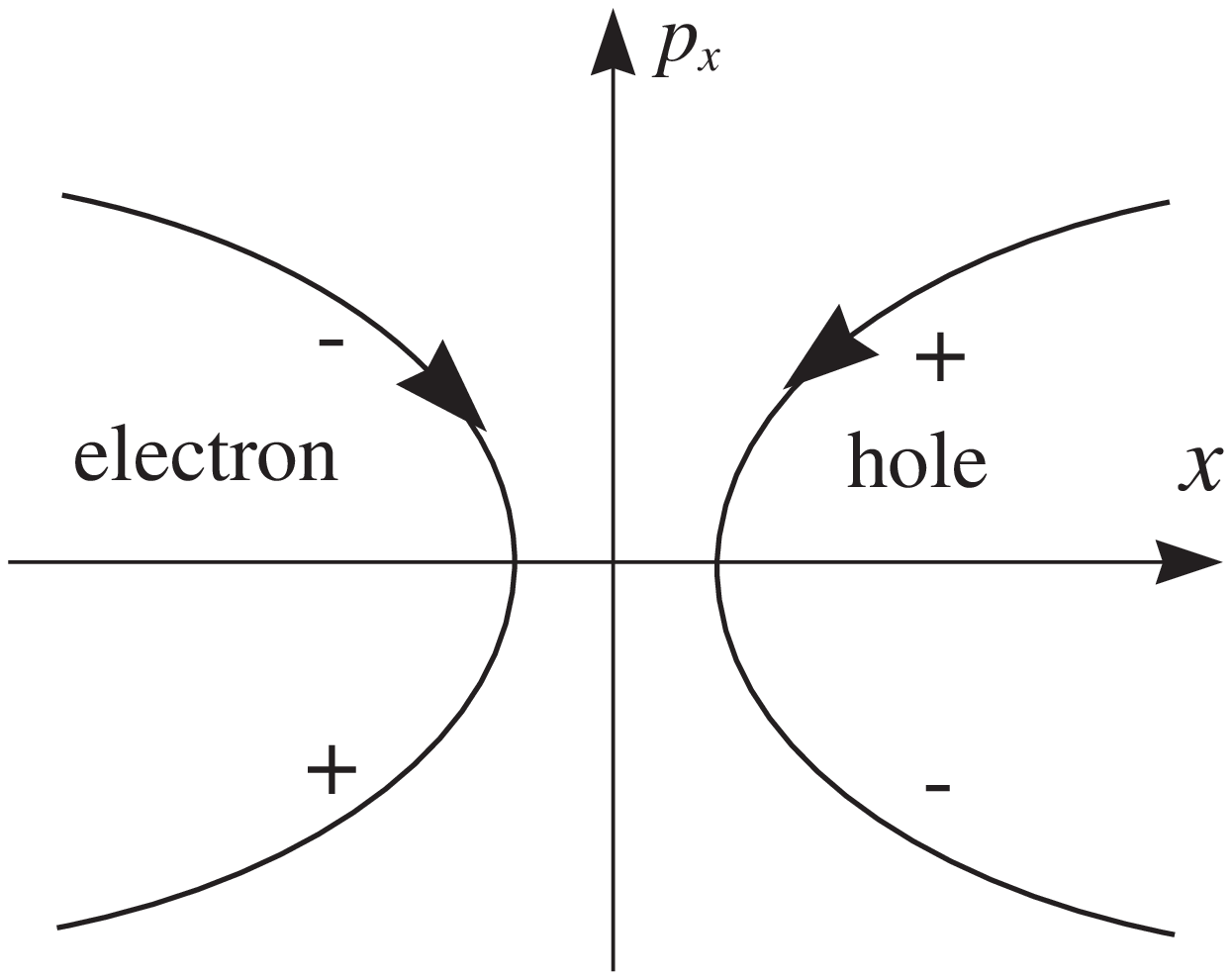} \\ \centerline{a)}}
\parbox{5cm}{\includegraphics[height=4cm]{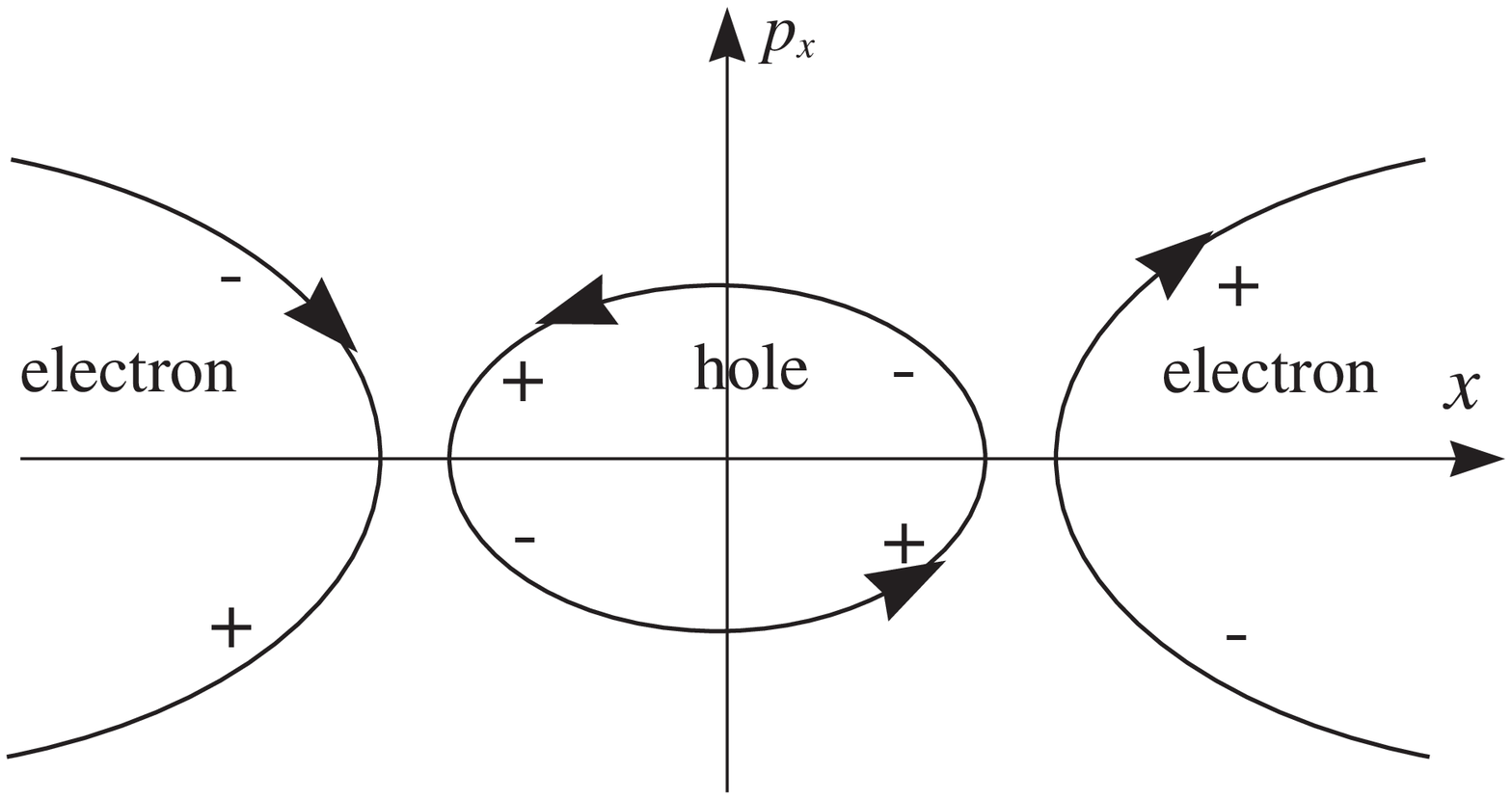} \\ \centerline{b)}}
\end{center}
\caption{Classical phase space for \textit{a}) n-p and \textit{b}) n-p-n junctions. In the electronic region the velocity is codirectional with the momentum and in the hole region the velocity has an opposite direction to the momentum. Therefore electronic trajectories are clockwise oriented but hole trajectories are oriented counterclockwise. Plus and minus in figures denote signs of $dp_x/dx$.}
\label{fig::phase-space}
\end{figure}

Let us first consider a scattering problem for $p_y\neq 0$ following \ref{app::sem-appr} and \ref{app::bohr-somm}. For an electron coming from the left of the classically forbidden region we have
\begin{eqnarray}
\fl \psi(x)&=&\frac{|E-u(x)|^{1/2}}{\left[(E-u(x))^2-p_y^2\right]^{1/4}}%\nonumber\\&\times&
\left(e^{iS^+(x)/h+i\phi^+_p(x)/2}+e^{-iS^+(x)/h+i\phi^-_p(x)/2-i\pi/2}\right)
\label{eq::amp-single}
\end{eqnarray}
and
\begin{eqnarray}
\fl\Psi(x)&=&\frac{|E-u(x)|^{1/2}}{\left[(E-u(x))^2-p_y^2\right]^{1/4}}%\nonumber\\&\times&
\left(\begin{array}{c}e^{iS^+(x)/h-i\phi^+_p(x)/2}+e^{-iS^+(x)/h-i\phi^-_p(x)/2-i\pi/2} \\
e^{iS^+(x)/h+i\phi^+_p(x)/2} + e^{-iS^+(x)/h+i\phi^-_p(x)/2-i\pi/2}
\end{array}\right),
\label{eq::Psi}
\end{eqnarray}
%
%$p_y>0$
%
%\begin{eqnarray}
%\Psi(x)&=&\frac{(E-u(x))^{1/2}}{\left|(E-u(x))^2-p_y^2\right|^{1/4}}
%\left(\begin{array}{c}2i\sin\bigl(S^+(x)/h-\phi^+_p(x)/2\bigr) \\
%\pm 2\cos\bigl(S^+(x)/h-\phi^+_p(x)/2\bigr)
%\end{array}\right)
%\end{eqnarray}
%
%$p_y<0$
%
%\begin{eqnarray}
%\Psi(x)&=&\frac{(E-u(x))^{1/2}}{\left|(E-u(x))^2-p_y^2\right|^{1/4}}
%\left(\begin{array}{c}2\cos\bigl(S^+(x)/h-\phi^+_p(x)/2\bigr) \\
%\pm 2i\sin\bigl(S^+(x)/h-\phi^+_p(x)/2\bigr)
%\end{array}\right)
%\end{eqnarray}
%
%\begin{equation}
% r(p_y)=e^{i\,\mathrm{sgn}(p_y)\pi/2-i\pi/2}=\mathrm{sgn}(p_y).
%\end{equation}
%
%\begin{equation}
% S_+'=\sqrt{v^2(x)-p_y^2}, \quad S_+''=v'(x)v(x)/\sqrt{v^2(x)-p_y^2}
%\end{equation}
%
where
\begin{eqnarray}
\fl S^\pm(x)=\pm\int_{x_0}^x \sqrt{v^2(x')-p_y^2}dx', \quad \phi^\pm_p(x)=\textrm{Arg}\,\left(\pm \sqrt{v^2(x)-p_y^2}+ip_y\right),\\
\phi_p^-(x)=\pi\mathrm{sgn}(p_y)-\phi_p^+(x), \quad v(x)=u(x)-E.
\end{eqnarray}
Note that $\phi_p^+(x)$ continuously depends on $p_y$ when it passes through zero and $\phi_p^-(x)$ undergoes a jump of $2\pi$.
The reflection coefficient $r$ can be computed from (\ref{eq::amp-single}) or (\ref{eq::Psi}). It is usually defined as the coefficient in front of the semiclassical solution corresponding to the outgoing wave. One can also assume that the potential tends to a constant at infinity and take the coefficient in front of the plane wave, which is a particular case of the definition given above. Obviously, the reflection coefficient defined in such a way does not depend on $x$. Choosing (\ref{eq::amp-single-pm}) as incoming and outgoing solutions, we can write wavefunctions on the left of the classically forbidden region as
\begin{eqnarray}
\fl \psi(x)=\frac{|E-u(x)|^{1/2}}{\left[(E-u(x))^2-p_y^2\right]^{1/4}}%\nonumber\\&\times&
\left(e^{iS^+(x)/h+i\phi^+_p(x)/2}+r(p_y)e^{-iS^+(x)/h+i\phi^-_p(x)/2}\right),
\label{eq::psi-refl}\\
\fl\Psi(x)=\frac{|E-u(x)|^{1/2}}{\left[(E-u(x))^2-p_y^2\right]^{1/4}}\nonumber\\
\times
\left[\left(\begin{array}{c}e^{iS^+(x)/h-i\phi^+_p(x)/2} \\
e^{iS^+(x)/h+i\phi^+_p(x)/2}
\end{array}\right)+r(p_y)
\left(\begin{array}{c}e^{-iS^+(x)/h-i\phi^-_p(x)/2} \\
e^{-iS^+(x)/h+i\phi^-_p(x)/2}
\end{array}\right)\right],
\label{eq::Psi-refl}
\end{eqnarray}
Comparing (\ref{eq::psi-refl}), (\ref{eq::Psi-refl}) and (\ref{eq::amp-single}), (\ref{eq::Psi}) we conclude that
%Using the identity
%\begin{equation}
% \phi_p^-(x)=\pi\mathrm{sgn}(p_y)-\phi_p^+(x)
%\end{equation}
%we can rewrite it in the form
%\begin{eqnarray}
%\fl \psi(x)&=&\frac{(E-u(x))^{1/2}}{\left|(E-u(x))^2-p_y^2\right|^{1/4}}%\nonumber\\&\times&
%\left(e^{iS^+(x)/h+i\phi^+_p(x)/2}+r(p_y)e^{-iS^+(x)/h-i\phi^+_p(x)/2}\right),
%\label{eq::amp-single2}
%\end{eqnarray}
\begin{equation}
 r(p_y)=e^{-i\pi/2}.
\label{eq::refl-np}
\end{equation}

A similar calculation for a hole coming from the right gives, see also figure \ref{fig::phase-space},
\begin{equation}
 r(p_y)=e^{i\pi/2}.
\label{eq::refl-pn}
\end{equation}
We paid attention to the definition of the reflection coefficient, since it may lead to discrepancy for the Dirac particle. The problem appears due to a jump of $2\pi$ in $\phi_p^-(x)$ at any fixed $x$ as a function of $p_y$ when it goes through zero. This jump is a consequence of the cut at $\phi_p=\pm \pi$. At any $p_y\neq 0$ this cut corresponds to infinite negative $x$-component of the momentum, and does not imply any discontinuities in the region, where the potential is finite. This jump results in the jump of $\pi$ in the phase of the wavefunction corresponding to the outgoing wave. However, the phase difference $\phi_p^-(x)/2-\phi_p^+(x)/2=\pi\,\mathrm{sgn}\,(p_y)/2-\phi_p^+(x)$ tends to zero when $x$ tends to a turning point $x_0$ and can therefore be treated as one half of the angle around the origin in $p$-space accumulating during the motion of a classicle particle from the point $x$ to the turning point $x_0$ and back. The peculiar behaviour of the phase difference can mathematicaly be expressed as the noncommutativity of limits:
\begin{eqnarray}
 \lim_{p_y\to \pm 0}\lim_{x\to x_0}[\phi_p^-(x)-\phi_p^+(x)]=0,\nonumber\\
  \lim_{x\to x_0}\lim_{p_y\to \pm0}[\phi_p^-(x)-\phi_p^+(x)]=\pm\pi.
\end{eqnarray}
The jump in the sign of the outgoing wave must be compensated by a kink in the reflection coefficient, since the whole wavefunction should analytically depend on $p_y$. To get rid of the jump one can redefine the outgoing wave and write \cite{Shytov08}
\begin{eqnarray}
\fl
\psi(x)&=&\frac{|E-u(x)|^{1/2}}{\left[(E-u(x))^2-p_y^2\right]^{1/4}}%\nonumber\\&\times&
\left(e^{iS^+(x)/h+i\phi^+_p(x)/2}+r(p_y)e^{-iS^+(x)/h-i\phi^+_p(x)/2}\right).
\end{eqnarray}
Though preserving the analyticity of $r(p_y)$, such a definition introduces an artificial jump of the phase as a function of $x$ upon reflection at negative $p_y$. Therefore we do not use it below.

The reflection coefficient, defined in accordance with (\ref{eq::psi-refl}), (\ref{eq::Psi-refl}) does not depend on the sign of $p_y$. It is completely defined by the orientation of the phase space, which is clockwise for an electron region and counterclockwise for a hole region (see figure \ref{fig::phase-space}). Finally, the reflection can be written as
\begin{equation}
 r(p_y)=e^{\mp i\pi/2},
\label{eq::refl-total}
\end{equation}
where `-' corresponds to electron and `+' to hole regions.

\begin{figure}
\begin{center}
\parbox{5cm}{\includegraphics[width=5cm]{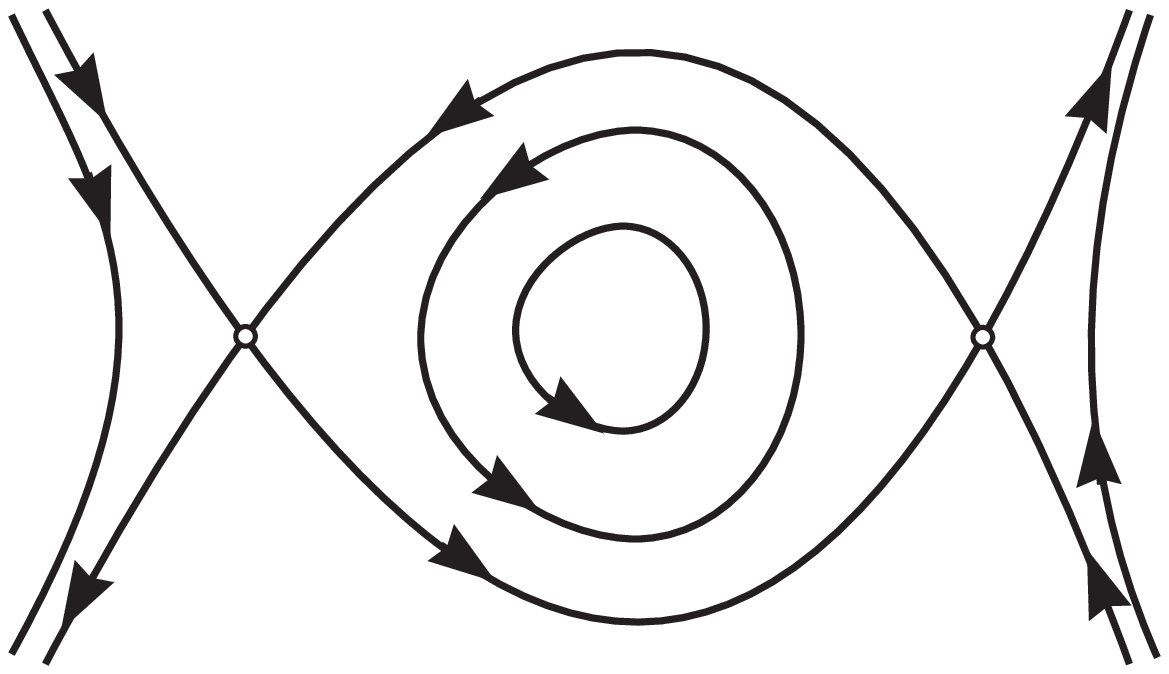} \\ \centerline{a)}}\hspace{.5cm}
\parbox{5cm}{\includegraphics[width=5cm]{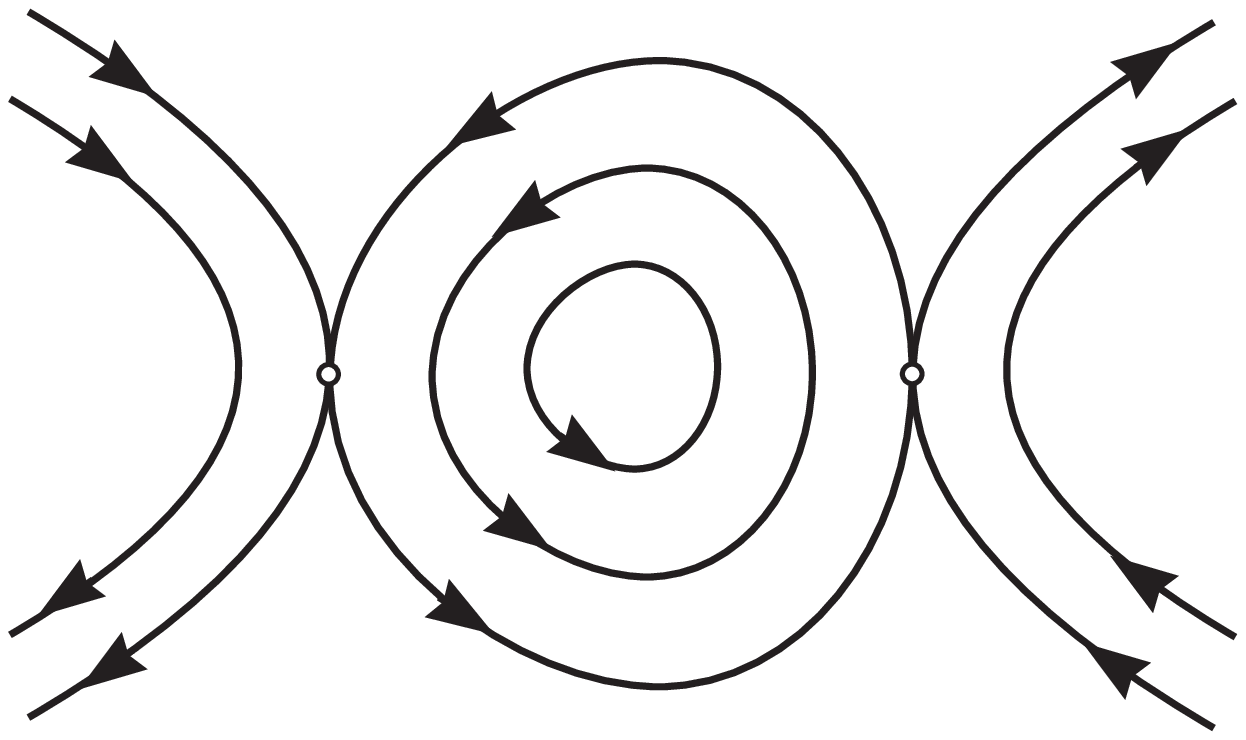} \\ \centerline{b)}}
\end{center}
\caption{Classical phase space for n-p-n junction in \textit{a}) single and \textit{b}) bilayer graphene. Different trajectories correspond to different values of $p_y$. One sees that for normal incidence (separatrices) the smoothest classical trajectory corresponds to total transmission in single layer graphene and to total reflection in bilayer graphene.}
\label{fig::phase-space-full}
\end{figure}

It is important to note, that the phase $-\pi/2$ and the module $1$ of the reflection coefficient (\ref{eq::refl-total}) were obtained under the assumption that there is no multiplicity change! It is not the case when $p_y\to 0$ and the trajectory in the phase space tends to a separatrix, see figure \ref{fig::phase-space-full}.
%\begin{equation}
% \phi^+(\infty)=0!!!
%\end{equation}

Let us now turn to bilayer graphene. The Hamiltonian describing the charge carrier dynamics reads
\begin{equation}
 H=\left(\begin{array}{cc}0 & (\hat p_x-ip_y)^2\\ (\hat p_x+ip_y)^2 & 0 \end{array}\right)+u(x)
\end{equation}
Eigenvalues and eigenvectors of $H(p_x,x)$ are
\begin{equation}
 L_0^\pm=\pm p^2+u(x), \qquad \chi_0^\pm=\frac{1}{\sqrt{2}}\left(\begin{array}{c}e^{-2i\phi_p} \\ \pm 1\end{array}\right).
\label{eq::LObi}
\end{equation}
We obtain
\begin{eqnarray}
\psi(x)=\frac{1}{\left|E-u(x)\mp p_y^2\right|^{1/4}}e^{\pm i S^+(x)/h+i\phi_p^\pm(x)},\nonumber\\
S(x)=\int_{x_0}^x\sqrt{\pm[E-u(x)]-p_y^2}dx.
\label{eq::psi-bilayer}
\end{eqnarray}

Obviously, the result (\ref{eq::refl-total}) is valid for the bilayer as well, since the orientation of the phase space is the same.

Between two classically forbidden regions effective Hamiltonians superimpose the following quantization conditions (see \ref{app::bohr-somm} for details):
\begin{equation}
 \frac{1}{h}\oint p_x dx + \frac{\beta}{2}\Delta\phi_p=2\pi\left(n+\frac{\nu}{4}\right),
\label{eq::quantization}
\end{equation}
where $\beta=1,\,2$ for single and bilayer respectively, $\nu=2$ is the Maslov index and $\Delta\phi_p$ is the total phase gain along the closed classical trajectory. The term $\beta\Delta\phi_p/2$ is the Berry phase in graphene \cite{Carmier08}. It is clear that $\Delta\phi_p$ acquires a non-zero value only if the trajectory in $p$-space encloses the origin. Therefore in the absence of magnetic field $\Delta\phi_p=0$. Quantization condition (\ref{eq::quantization}) allows one to determine resonance angles.

Though the considered diagonalization is very powerful to deal with complicated matrix Hamiltonians in a classically allowed region, it possesses a substantial disadvantage: it treats electrons and holes separately neglecting tunneling effects. In the classically forbidden region when $|p|=0$, i.e. $p_x=ip_y$ effective Hamiltonians $L_0^\pm$ become degenerate. At this point electron to hole transition may occur and the diagonalization fails. This transition is the origin of the Klein tunneling.

\section{\label{sec::normal}Normal incidence}

In the case of normal incidence $p_y=0$ and at the point $x_0$, where $u(x_0)=E$ there is a multiplicity change, i.e. effective Hamiltonians $L_0^\pm$ become degenerate (see figure \ref{fig::phase-space-full}). To study wavefunctions in this case one can not apply a standard semiclassical treatment, described in Section \ref{sec::standard-semi}, since there may be a ``jump'' between $L^+$ and $L^-$. Fortunately, for the normal incidence in graphene there is an exact pseudospin conservation, which allows one to study this case in detail.

For $p_y=0$ equations (\ref{eq::single-dimless}), (\ref{eq::bi-dimless}) read
\begin{equation}
 [\sigma_x \hat p_x^\beta + u(x)-E]\Psi=0,
 \label{eq::normal-single-bi}
\end{equation}
where $\beta=1,\,2$ for single and bilayer respectively. Eigenvectors of $\sigma_x \hat p_x^\beta$ do not depend on $\hat p_x^\beta$, therefore (\ref{eq::normal-single-bi}) can easily be diagonalized, which leads to
\begin{equation}
 [\pm \hat p_x^\beta + u(x)-E]\eta_{1,2}=0,
\label{eq::normal}
\end{equation}
where
\begin{equation}
\Psi=
\left(\begin{array}{c}1 \\ 1\end{array}\right)\eta_1+\left(\begin{array}{c}1 \\ -1\end{array}\right)\eta_2.
\end{equation}
In this case the eigenvalue of $\sigma_x$ (``pseudospin'') persists. 
Pseudospin conservation leads to very different physical consequences for single and bilayer graphene.

For single layer graphene pseudospin conservation means the conservation of the $x$-component of the velocity. 
Equation (\ref{eq::normal}) is the first order differential equation, which can be solved exacly. We obtain
\begin{equation}
 \eta_{1,2}=C_{1,2}\exp\left(\pm i\int_{x_0}^x[E-u(x')]dx'\right),
\label{eq::exact-single}
\end{equation}
where $C_{1,2}$ are some constants. The absence of the reflected wave in (\ref{eq::exact-single}) means that for any potential shape one has a perfect transmission. Thus we conclude that at the point $p_x=0$ there is a \textit{total transition} between electron and hole states since Hamiltonians (\ref{eq::L0-single}) depend on $|p|$ in contrast to (\ref{eq::normal})!

For bilayer graphene pseudospin conservation is equivalent to the conservation of particle type, as is seen from the comparison of (\ref{eq::LObi}) and (\ref{eq::normal}). Therefore, an incoming particle obeys the Schr\"odinger equation (\ref{eq::normal}) everywhere.  For a ``Klein-setup'' this leads to exponentially decaying transmission as a function of a potential width and height.

Total transmission for normally incident electrons in single layer graphene and its exponential damped behaviour in bilayer have natural explanations of classical phase space (figure \ref{fig::phase-space-full}). In both cases the most probable process corresponds to the smoothest trajectory, constructed from separatrix pieces. For the single layer such a trajectory goes through the barrier and gives total transmission, while for bilayer one has to choose the trajectory reflected from the barrier to avoid discontinuity in the second derivative.

\section{Exact reduction to effective Schr\"odinger equations}

\begin{figure}
\begin{center}
\includegraphics[width=10cm]{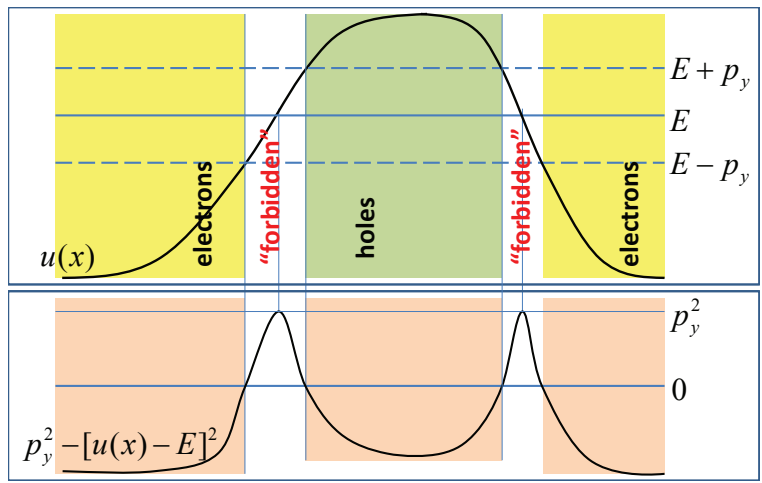}
\end{center}
\caption{Comparison of the initial potential and the real part of the effective potential.}
\label{fig::reduction}
\end{figure}

In Section \ref{sec::standard-semi} it was discussed that the standard adiabatic diagonalization fails to describe Klein tunneling, since it treats electrons and holes separately. However the existence of the exact diagonalization (Section \ref{sec::normal}) for a normal incidence raises the issue of a possible generalization for angular scattering. We shall show that for a single layer graphene there exists an \textit{exact} transformation, reducing the original Dirac equation to a scalar Schr\"odinger-like equation with a \textit{complex} potential. It is clear that such a procedure can not be a unitary transformation of the original Hermitian Hamiltonian.

Let us turn back to the Dirac equation for single layer and write it in the form
\begin{equation}
 \bigl(\boldsymbol{\sigma} \boldsymbol{p}+v(x)\bigr)\Psi=0,
\label{eq::sigma-single}
\end{equation}
where $\boldsymbol{p}=(\hat p_x,p_y)$, $v(x)=u(x)-E$. Let us act on the last equation from the left by the operator $\boldsymbol{\sigma} \boldsymbol{p}-v(x)$. Then we get
\begin{eqnarray}
 \bigl(\boldsymbol{\sigma} \boldsymbol{p}-v(x)\bigr)\bigl(\boldsymbol{\sigma} \boldsymbol{p}+v(x)\bigr)\Psi =
\bigl(\hat p_x^2 + p_y^2 - v(x)^2 + \sigma_x [\hat p_x,v(x)]\bigr)\Psi\nonumber\\
\qquad=\bigl(\hat p_x^2 + p_y^2 - v(x)^2 - ih\sigma_x v'(x)\bigr)\Psi=0.
\label{eq::red1}
\end{eqnarray}
Remarkably, (\ref{eq::red1}) contains only the single matrix $\sigma_x$. Therefore it can easily be diagonalized. We write
\begin{equation}
\Psi=
\left(\begin{array}{c}1 \\ 1\end{array}\right)\eta_1+\left(\begin{array}{c}1 \\ -1\end{array}\right)\eta_2
\label{eq::Psi-eta}
\end{equation}
and obtain
\begin{equation}
 \left(h^2\frac{d^2}{dx^2}+v(x)^2-p_y^2\pm ihv'(x)\right)\eta_{1,2}=0.
\label{eq::reduction}
\end{equation}
Functions $\eta_{1,2}$ are not independent and the connection formula can be obtained from (\ref{eq::sigma-single}). Function $\eta_2$ can be reconstructed from $\eta_1$ as
\begin{equation}
 \eta_2=\frac{1}{p_y}\left(h\frac{d}{dx}+iv(x)\right)\eta_1.
\label{eq::connection}
\end{equation}

In figure \ref{fig::reduction} one sees the initial potential landscape and the real part of the effective potential in (\ref{eq::reduction}).
Though this equation takes the form of a Schr\"odinger equation, there are two substantial distinctions as compared to a common Schr\"odinger particle: \textit{i}) the effective potential is complex and \textit{ii}) it depends on energy.

\section{Single n-p junction}

\subsection{Exact solution in the case of a linear n-p junction}

Let us first consider an exacly solvable model for a linear potential $v(x)=\alpha x$, $\alpha>0$ \cite{Cheianov06}. Introducing a new variable $x'=(\alpha/h)^{1/2}x$ and new $y$-component of the momentum $p_y'=(h\alpha)^{-1/2}p_y$ we exclude $h$ and $\alpha$ from (\ref{eq::reduction}). Then it takes the form (we omit primes):
\begin{equation}
 \left(\frac{d^2}{dx^2}+x^2-p_y^2+i\right)\eta_1=0.
\label{eq::etalon}
\end{equation}
%In what follows this equation plays a role of an ``etalon-equation'' \cite{Berry72} for Dirac particles.
Introducing a new variable $z$ such that $x=\xi z$ we have:
\begin{equation}
 \left(\frac{d^2}{dz^2}+\xi^2(i-p_y^2)+\xi^4 z^2\right)\eta_1=0.
\label{eq::Weber0}
\end{equation}
Choosing an appropriate value for $\xi$ we can reduce (\ref{eq::Weber0}) to the Weber's equation \cite{Abramowitz65}
\begin{equation}
 w''(z)+\left(\nu+\frac{1}{2}-\frac{z^2}{4}\right)w(z)=0.
\end{equation}

Indeed, choosing $\xi=e^{-i\pi/4}/\sqrt{2}$ and solving the Weber's equation \cite{Abramowitz65,WolframFunctions} we obtain
\begin{equation}
 \eta_1=c_1 D_\nu(\sqrt{2}e^{i\pi/4}x)+c_2 D_{-\nu-1}(\sqrt{2}e^{3i\pi/4}x),
\label{eq::eta1}
\end{equation}
where $\nu=ip_y^2/2$ and $D_\nu$ are the parabolic cylinder functions. For these functions the following identities hold:
\begin{eqnarray}
 \frac{\partial D_\nu(z)}{\partial z}=\nu D_{\nu-1}(z)-\frac{z}{2}D_\nu(z), \quad %\nonumber\\
 \frac{\partial D_\nu(z)}{\partial z}=\frac{z}{2}D_\nu(z)-D_{\nu+1}(z).
\label{eq::pcyl-id}
\end{eqnarray}
Applying the first equality from (\ref{eq::pcyl-id}) we find
\begin{eqnarray}
 \left(\frac{\partial}{\partial x}+ix\right)D_\nu(\sqrt{2}e^{i\pi/4}x)=\sqrt{2}\nu e^{i\pi/4}D_{\nu-1}(\sqrt{2}e^{i\pi/4}x).
\label{eq::par-cyl1}
\end{eqnarray}
From the second equality in (\ref{eq::pcyl-id}) we have
\begin{eqnarray}
 \left(\frac{\partial}{\partial x}+ix\right)D_{-\nu-1}(\sqrt{2}e^{3i\pi/4}x)=\sqrt{2} e^{-i\pi/4}D_{-\nu}(\sqrt{2}e^{3i\pi/4}x).
\label{eq::par-cyl2}
\end{eqnarray}
Substituting (\ref{eq::par-cyl1}), (\ref{eq::par-cyl2}) into (\ref{eq::connection}) we obtain:
\begin{eqnarray}
 \eta_2 &=& \frac{1}{p_y}\left(\frac{d}{dx}+ix\right)\eta_1\nonumber\\
&=&\frac{c_1}{p_y}\sqrt{2}\nu e^{i\pi/4}D_{\nu-1}(\sqrt{2}e^{i\pi/4}x)+\frac{c_2}{p_y}\sqrt{2} e^{-i\pi/4}D_{-\nu}(\sqrt{2}e^{3i\pi/4}x).
\label{eq::eta2}
\end{eqnarray}
From (\ref{eq::eta1}), (\ref{eq::eta2}) we have for $\Psi=(\psi_1,\psi_2)$:
\begin{eqnarray}
 \psi_{1,2}(x)&=&\eta_1(x)\pm\eta_2(x)\nonumber\\
&=&c_1\left(D_\nu(\sqrt{2}e^{i\pi/4}x)\pm
\frac{\sqrt{2}\nu e^{i\pi/4}}{p_y}D_{\nu-1}(\sqrt{2}e^{i\pi/4}x)\right)
\nonumber\\
&+&c_2\left(D_{-\nu-1}(\sqrt{2}e^{3i\pi/4}x)\pm
\frac{\sqrt{2} e^{-i\pi/4}}{p_y}D_{-\nu}(\sqrt{2}e^{3i\pi/4}x)\right).
\end{eqnarray}

Using the asymptotic expansions of the parabolic cylinder functions (see \ref{app::parcas}), we find when $x\to\infty$ (hole region):
\begin{eqnarray}
\psi_{1,2} &\to& c_1 z_1^\nu e^{-ix^2/2} \nonumber\\
&+&c_2\left[-\frac{\sqrt{2\pi}}{\Gamma(\nu+1)}z_2^\nu e^{-ix^2/2-i\pi(\nu+1)}\pm\frac{\sqrt{2}e^{-i\pi/4}}{p_y}z_2^{-\nu} e^{ix^2/2}\right],
\end{eqnarray}
and when $x\to-\infty$ (electron region):
\begin{eqnarray}
\psi_{1,2} &\to&
c_1\left[(\bar z_2)^\nu e^{-ix^2/2}\pm\frac{\sqrt{2\pi}}{\Gamma(1-\nu)}\frac{\sqrt{2}\nu e^{i\pi/4}}{p_y}(\bar z_2)^{-\nu} e^{ix^2/2-i\pi\nu}\right]\nonumber\\
&\pm&c_2\frac{\sqrt{2}e^{-i\pi/4}}{p_y} (\bar z_1)^{-\nu} e^{ix^2/2},
\end{eqnarray}
where $z_1=\sqrt{2}e^{i\pi/4}|x|$, $z_2=\sqrt{2}e^{3i\pi/4}|x|$ and a bar means complex conjugation.

Now we turn to the discussion of the scattering problem. While tunneling through the barrier the Dirac particle turns from an electron to a hole or vice versa. The \mbox{$x$-component} of the group velocity of the hole $v_x=\partial L^-_0/\partial p_x=-p_x/p$ has an opposite sign with respect to its momentum $p_x$. Let us consider an electron, coming from $-\infty$ with a positive velocity $v_x$. It corresponds to the action $S^+(x)\simeq -x^2/2$, since $p_x=\partial S^+(x)/\partial x\simeq-x>0$ and $v_x=p_x/p>0$. Thus, the reflected electron corresponds to $S^-(x)\simeq x^2/2$. The transmitted hole with a positive velocity has a negative momentum $p_x$. Hence it corresponds to the action $S^-(x)\simeq -x^2/2$. From the absence of the incoming wave in the hole region we find $c_2=0$.

Let us consider (\ref{eq::Psi-refl}) at infinity. Then we have for the action:
\begin{eqnarray}
S^+(x)&=&\int_{\mathrm{sgn}\,(x)|p_y|}^x\sqrt{y^2-p_y^2}dy\nonumber\\
&=&\frac{1}{2}\mathrm{sgn}\,(x)\left\{|x|\sqrt{x^2-p_y^2}-p_y^2
\ln\left[\left|\frac{x}{p_y}\right|+\sqrt{\left(\frac{x}{p_y}\right)^2-1}\right]\right\},
\end{eqnarray}
where we assumed that $|x|>|p_y|$. For large $x$ we obtain
\begin{equation}
 S^+(x)\simeq \frac{1}{2}\mathrm{sgn}\,(x)\left\{x^2-\frac{p_y^2}{2}-p_y^2
\ln\left(\frac{2|x|}{|p_y|}\right)\right\}.
\end{equation}
Thus for large negative $x$ (\ref{eq::Psi-refl}) reads
\begin{eqnarray}
\fl \Psi(x)=%\nonumber\\&\times&
e^{-ix^2/2+ip_y^2/4+(i/2)p_y^2\ln(2|x|/|p_y|)}\left(\begin{array}{c}1 \\ 1\end{array}\right)\nonumber\\
+r(p_y)e^{ix^2/2-ip_y^2/4-(i/2)p_y^2\ln(2|x|/|p_y|)}
\left(\begin{array}{c}e^{-i\pi\,\mathrm{sgn}\,(p_y)/2} \\ e^{i\pi\,\mathrm{sgn}\,(p_y)/2}\end{array}\right).
\label{eq::amp-single-inf}
\end{eqnarray}
This gives for the reflection
\begin{equation}
 r(p_y)=\frac{\sqrt{\pi}|p_y|}{\Gamma(1-\nu)}e^{-\pi p_y^2/4}e^{i\theta(p_y)-i\pi/2},
\end{equation}
where $\theta(p_y)=p_y^2/2-(p_y^2/2)\ln(p_y^2/2)-\pi/4$. Using equalities
\begin{eqnarray}
 |\Gamma(1-\nu)|^2=\Gamma(1-\nu)\Gamma(1+\nu)=\nu\Gamma(\nu)\Gamma(1-\nu)=\frac{\pi\nu}{\sin(\pi\nu)}\nonumber\\
=\frac{\pi p_y^2}{e^{\pi p_y^2/2}-e^{-\pi p_y^2/2}}
\end{eqnarray}
we can write the reflection coefficient as
\begin{equation}
 r(p_y)=\sqrt{1-e^{-\pi p_y^2}}e^{i\theta(p_y)-i\gamma(p_y)-i\pi/2},
\end{equation}
where $\gamma(p_y)=\mathrm{Arg}\,\Gamma(1-ip_y^2/2)$. From the asymptotic expansion of the $\Gamma$-function at large arguments \cite{Abramowitz65}, one concludes that $\gamma(p_y)$ tends to $\theta(p_y)$ when $p_y$ tends to infinity. At small $p_y$ the reflection coefficient is proportional to $|p_y|$. This nonanalytic behaviour is compensated by the jump of the phase of the reflected wave as we discussed in Section~\ref{sec::standard-semi}.

Comparing the coefficient in front of incoming and transmitted waves we find for the transmission amplitude
\begin{equation}
 t=e^{i\pi\mathrm{sgn}\,(p_y)/2}e^{i\pi\nu}=e^{i\pi\mathrm{sgn}\,(p_y)/2}e^{-\pi p_y^2/2}.
\label{eq::tranmission}
\end{equation}
For the transmission probability we thus have
\begin{equation}
 |t|^2=e^{-\pi p_y^2}=e^{-\pi p^2\sin^2\phi_p}
\end{equation}
This result was first obtained by Cheianov and Fal'ko \cite{Cheianov06}. Considering the scattering from the right to the left we find that the transmission amplitude in this case is
\begin{equation}
 t=-e^{-i\pi\mathrm{sgn}\,(p_y)/2}e^{-\pi p_y^2/2}.
\label{eq::tranmission2}
\end{equation}

The transfer matrix connecting incoming and outgoing waves from the right to the left of the barrier for positive $\alpha$ is
\begin{eqnarray}
\fl
T_+=e^{-i\pi\mathrm{sgn}\,(p_y)/2}\left(\begin{array}{cc}e^{\pi p_y^2/2} & \displaystyle{\left(e^{\pi p_y^2}-1\right)^{1/2}e^{i(\gamma-\theta-\pi/2)}} \\
\displaystyle{\left(e^{\pi p_y^2}-1\right)^{1/2}e^{i(\theta-\gamma-\pi/2)}} & -e^{\pi p_y^2/2} \end{array}\right).
\label{eq::transfer-linpot}
\end{eqnarray}
For negative $\alpha$ the transfer matrix reads
\begin{eqnarray}
\fl
T_-=e^{i\pi\mathrm{sgn}\,(p_y)/2}\left(\begin{array}{cc}-e^{\pi p_y^2/2} & \displaystyle{\left(e^{\pi p_y^2}-1\right)^{1/2}e^{i(\theta-\gamma-\pi/2)}} \\
\displaystyle{\left(e^{\pi p_y^2}-1\right)^{1/2}e^{i(\gamma-\theta-\pi/2)}} & e^{\pi p_y^2/2} \end{array}\right).
\end{eqnarray}

\subsection{Transmission probability in semiclasical approximation}

Let us consider an outgoing hole on the right of a generic potential monotonously growing from the left to the right. In semiclassical approximation it is described by the wavefunction
 \begin{eqnarray}
\Psi(x)=\frac{e^{-i\pi\,\mathrm{sgn}\,(p_y)/2}|E-u(x)|^{1/2}}{\left[(E-u(x))^2-p_y^2\right]^{1/4}}e^{-iS^+(x)/h}
\left(\begin{array}{c}e^{i\phi^+_p(x)/2} \\
e^{-i\phi^+_p(x)/2}
\end{array}\right).
\label{eq::Psi-out}
\end{eqnarray}
Using equalities
\begin{eqnarray}
 \cos\left(\frac{\phi_p^+}{2}\right)=\sqrt{\frac{1+\cos\phi_p^+}{2}}, \quad
\sin\left(\frac{\phi_p^+}{2}\right)=\sqrt{\frac{1-\cos\phi_p^+}{2}}, \nonumber\\
\cos\phi_p^+=\frac{|p_x|}{|p|}=\frac{[(E-u(x))^2-p_y^2]^{1/2}}{|E-u(x)|}
\end{eqnarray}
we write it as
 \begin{eqnarray}
\fl\Psi(x)=\frac{e^{-i\pi\,\mathrm{sgn}\,(p_y)/2}}{\left[v^2-p_y^2\right]^{1/4}\left[v+\sqrt{v^2-p_y^2}\right]^{1/2}}e^{-iS^+/h}
\left(\begin{array}{c}v+\sqrt{v^2-p_y^2}+ip_y \\
v+\sqrt{v^2-p_y^2}-ip_y
\end{array}\right).
\label{eq::Psi-out2}
\end{eqnarray}
with $v(x)=u(x)-E$. According to (\ref{eq::Psi-out2}) the components of $\Psi(x)$ can be represented \textit{exactly} as a sum or a difference of functions $\eta_1$, $\eta_2$ obeying Schr\"odinger-like equations (\ref{eq::reduction}) with a complex potential. Despite the complexity of the potential, according to \cite{Fedoruk66}, to connect a transmitted wave on the right of the barrier with an incoming wave on the left of the barrier one can still use an analytic continuation in the classically forbidden region similar to \cite{Landau77}. In contrast to a usual Schr\"odinger equation the transmitted hole has a negative momentum, so $\Psi(x)$ in (\ref{eq::Psi-out}) is proportional to $e^{-iS^+/h}$, but not to $e^{iS^+/h}$. Therefore the passage should be done in the lower complex half-plane. Since both functions $\eta_1$ and $\eta_2$ allow analytic continuations in the lower half-plane, $\Psi(x)$ can also be continued in the lower half-plane.

Performing the passage and connecting the outgoing wave with an incoming wave, we obtain for the transmission coefficient:
\begin{equation}
 t=e^{i\pi\mathrm{sgn}\,(p_y)/2}e^{-K/h}, \quad K=\left|\int_{x_1}^{x_2}\sqrt{p_y^2-v^2(x)}dx\right|,
\label{eq::tranmission-semi}
\end{equation}
where $x_1$ and $x_2$ are two turning points, i.e. solutions for the equation $p_y^2-v^2(x)=0$. The standard complex WKB technique \cite{Heading62,Froeman65,Fedoruk66} does not allow one to compute the reflection coefficient with an exponential accuracy. The module of the reflection coefficient can be computed from the unitarity of the scattering matrix: $|r|^2+|t|^2=1$. This gives
\begin{equation}
 |r|=\sqrt{1-e^{-2K/h}}.
\end{equation}
The semiclassical phase of the reflection coefficient can be reconstructed from (\ref{eq::refl-np}), (\ref{eq::refl-pn}). Thus in semiclassical approxiamtion we obtain:
\begin{equation}
 r=\sqrt{1-e^{-2K/h}}e^{\mp i\pi/2},
\end{equation}
where `-' corresponds to the electron and `+' to the hole region. For small $p_y$ any potential can be linearized in the classically forbidden region. Comparing (\ref{eq::tranmission}) and (\ref{eq::tranmission-semi}) we see that in the limit $p_y\to 0$ the semiclassical transmission becomes exact. Therefore (\ref{eq::tranmission-semi}) can be used as a uniform approximation for the transmission coefficient at any $p_y$. Then the uniform approximation for the reflection coefficient reads
\begin{equation}
 r=\sqrt{1-e^{-2K/h}}e^{\mp i\pi/2+i\Theta},
\end{equation}
where $\Theta$ tends to zero when $p_y$ tends to infinity. For the phase $\Theta$ we used the expression \cite{Berry72,Froeman02}
\begin{equation}
 \Theta=\frac{K}{\pi h}-\frac{K}{\pi h}\ln\left(\frac{K}{\pi h}\right)-\frac{\pi}{4}-\mathrm{Arg}\,\Gamma\left(1-\frac{iK}{\pi h}\right),
\end{equation}
which was obtained by the replacement $\pi p_y^2/2$ by $K/h$ in $\theta(p_y)-\gamma(p_y)$.

\section{Klein tunneling in n-p-n junctions. Fabry-P\'erot interferometer}

Let us consider some generic potential barrier $u(x)$. In graphene it implies n-p-n junction (see figure \ref{fig::reduction}). In terms of (\ref{eq::reduction}) n-p-n junction becomes a complex double-hump potential. The transfer matrix in this case is
\begin{eqnarray}
 T=T_{+} \left(\begin{array}{cc} e^{iS/h} & 0 \\ 0 & e^{-iS/h} \end{array}\right) T_{-},
\quad S=\int_{x_2}^{x_3} \sqrt{v^2(x)-p_y^2}dx,
\end{eqnarray}
where we assume that $x_2<x_3$. There is no extra phase coming from $\phi_p^\pm$ since each of these functions takes the same values at both turning points $x_2$, $x_3$, lying in the hole region. The transmission coefficient reads
\begin{eqnarray}
 t=\frac{1}{T_{11}}=-
\frac{e^{-iS/h}e^{-K_1/h-K_2/h}}{1+e^{-2iS/h-i\Theta_1-i\Theta_2}\sqrt{1-e^{-2K_1/h}}\sqrt{1-e^{-2K_2/h}}}. \label{eq:pnptransmission}
\end{eqnarray}
The obtained transmission amplitude can easily be treated in terms of a sum of probability amplitudes of multiscattering processes leading to transmission \cite{Shytov08}. One sees that for normal incidence $K_1 = K_2 = 0$ and the module of the transmission coefficient becomes one. Transmission resonances can be found from the condition:
\begin{equation}
 \frac{S}{h} + \frac{\Theta_1+\Theta_2}{2} = \pi\left(n+\frac{1}{2}\right),
\end{equation}
which coincides with the quantization condition (\ref{eq::quantization}) for large $p_y$. For a symmetric n-p-n junction the resonant transmission is always one, since $K_1 = K_2$. For an asymmetric junction resonant values of transmission decay as
\begin{equation}
 t \sim \frac{1}{\cosh(K_1/h-K_2/h)}
\label{eq::t-assym-res}
\end{equation}
when $K_1/h\gg 1$ and $K_2/h\gg 1$. From (\ref{eq::t-assym-res}) one sees that resonant values of transmission exponentially decay as a function of $|K_1-K_2|$. Such a fast decay can be crucial if one wants to weaken the influence of side resonances.

\section{Numerical results} \label{sec:numerics}
\begin{figure}[p]
\begin{center}
\includegraphics[height=7cm]{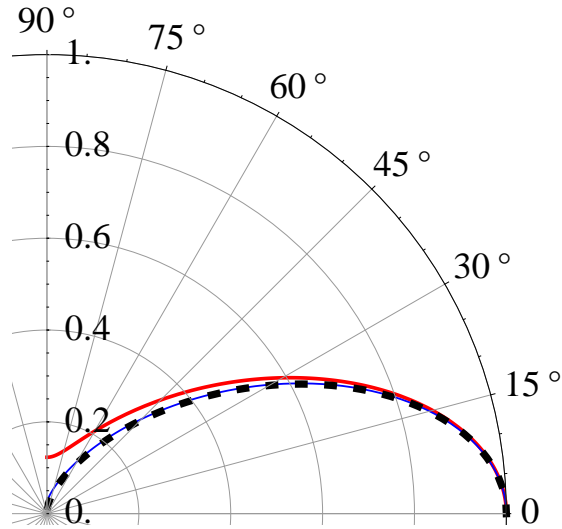}
\end{center}
\caption{The angular dependence of the transmission coefficient for a particle of energy 80 meV incident on an n-p junction of height 200 meV. The potential is given by equation~(\ref{eq:numericpotential}), with $l_1 = 70$ nm. The blue line shows the numerical result with 49 steps, the dashed line shows the semiclassical result~(\ref{eq::tranmission-semi}) and the red line shows the result for a linear potential~(\ref{eq::tranmission}), where the parameter $\alpha$ was taken as the derivative at the central point of the junction.}
\label{fig::single-hump}
\end{figure}

\begin{figure}[p]
\begin{center}
\includegraphics[height=7cm]{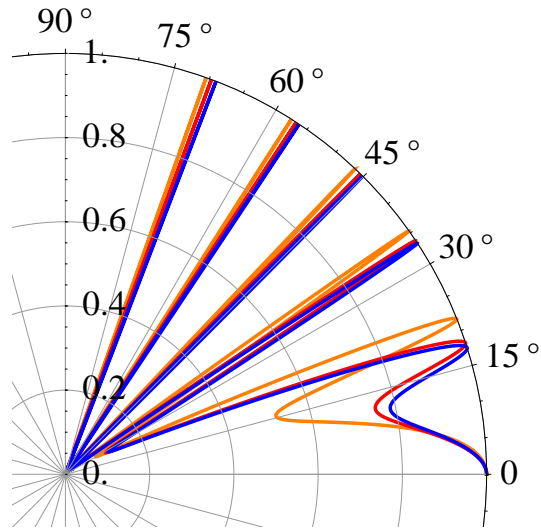}
\end{center}
\caption{The angular dependence of the transmission coefficient for a particle of energy 80 meV incident on an n-p-n junction of height 200 meV. The barrier width $l_2 = 250$~nm and n-p and p-n regions have characteristic lengths $l_1 = l_3 = 100$ nm. The blue line shows the numerical result for 99 steps, the red line shows the uniform approximation~(\ref{eq:pnptransmission}) and the orange line shows the semiclassical answer ($\Theta_1=\Theta_2=0$).}
\label{fig::double-hump}
\end{figure}

\begin{figure}[p]
\begin{center}
\includegraphics[height=7cm]{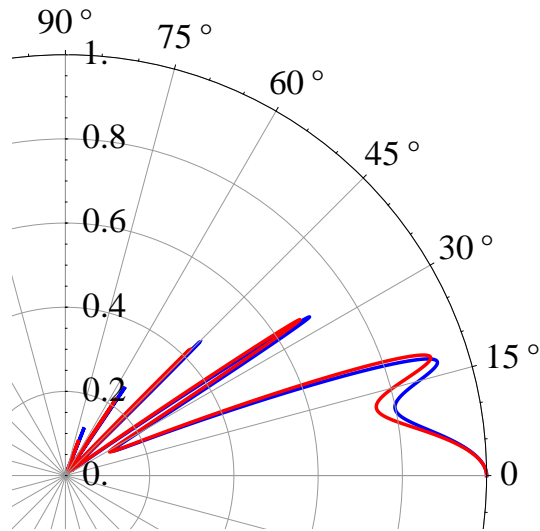}
\end{center}
\caption{The angular dependence of the transmission coefficient for a particle of energy 80 meV incident on an n-p-n junction of height 200 meV. The barrier width $l_2 = 250$~nm and the n-p and p-n regions have characteristic lengths $l_1 = 150$ nm and $l_3 = 50$~nm, respectively. The blue line shows the numerical results for 99 steps, while the red line shows the uniform approximation~(\ref{eq:pnptransmission}).}
\label{fig::single-asym}
\end{figure}

In figures~\ref{fig::single-hump}-\ref{fig::single-asym} we compare our semiclassical predictions with numerical results, obtained from a multistep approximation of the initial potential. A check on the accuracy of the calculation for constant $p_y$ is provided by the current \begin{equation}
   j_x = \Psi^\dagger \sigma_x \Psi, \quad d\,j_x/dx=0.\label{eq:wronskiansinglelayer}
\end{equation}

To simulate an n-p junction we used the potential
\begin{equation}
  V(x/l_1) = 0.5\,U_{max}\left[1 + \tanh(10x/l_1 - 5)\right] \\ \label{eq:numericpotential}
\end{equation}
with a characteristic length scale $l_1$. An n-p-n junction was simulated as an n-p junction with a characteristic length $l_1$, a p-n junction with a characteristic length $l_3$ and a constant potential in between of the length $l_2$.

In figure~\ref{fig::single-hump} we show the comparison of our numerical result for an n-p junction with the semiclassical transmission~(\ref{eq::tranmission-semi}) and the transmission for a linear potential~(\ref{eq::tranmission}). 
While the semiclassical prediction works uniformly over the entire range of angles, the prediction obtained from a linear potential works only for small angles.

In figure~\ref{fig::double-hump} we show the comparison of our numerical results with the prediction~(\ref{eq:pnptransmission}) for a symmetric n-p-n junction. We also show the semiclassical result, which is obtained by setting $\Theta_1=\Theta_2=0$. The agreement between the latter answer and numerics gets better as the angle increases, i.e. deep in the semiclassical regime. The result~(\ref{eq::tranmission-semi}) uniformly approximates the numerical data over the entire range of angles.

In figure~\ref{fig::single-asym} the result for an asymmetric n-p-n junction is shown. The height of resonances is seen to decay. The \textit{suppression of side resonances for asymmetric junctions} in single layer graphene can have essential consequences for attemps to confine Dirac particles!

Numerical computations for bilayer graphene using the above procedure are less accurate, due to the presence of real exponentials everywhere. Therefore we were unable to check the  quantization condition~(\ref{eq::quantization}) numerically with a high precision. To check the accuracy of the computation we used the current
\begin{eqnarray}
  j_x &=& \psi_1 \left( \frac{d}{d x} + k_y \right) \psi_2^* - \psi_2^* \left( \frac{d}{d x} - k_y \right) \psi_1 \\
   &+&
  \psi_2 \left( \frac{d}{d x} - k_y \right) \psi_1^* - \psi_1^* \left( \frac{d}{d x} + k_y \right) \psi_2 ,
\quad d\,j_x/dx=0. \label{eq:wronskianbilayer}
\end{eqnarray}

\begin{figure}[p]
\begin{center}
\includegraphics[height=7cm]{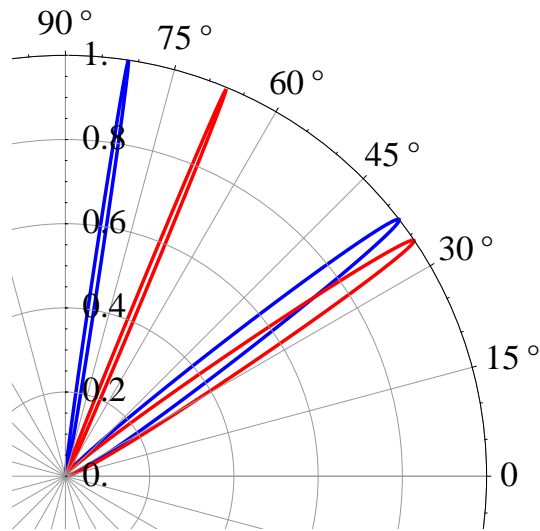}
\end{center}
\caption{The angular dependence of the transmission coefficent for a particle of energy 17 meV incident on symmetric and asymmetric n-p-n junctions in bilayer graphene. Each junction has a height of  50 meV and a width $l_2 = 100$ nm. The blue line shows the numerical result for a symmetric junction with $l_1 = l_3 = 10$ nm, while the red line shows an asymmetric junction with $l_1 = 20$ nm and $l_3 = 40$ nm. All calculations were done with 99 steps per junction.}
\label{fig::bilayer}
\end{figure}

In figure~\ref{fig::bilayer} we show our numerical results for a symmetric and an asymmetric n-p-n junction with the same shape of the potential as before. In contrast to the case of single layer, resonances \textit{do not seem to decay in this case}. 

\section{Conclusion}

Let us summarize our main results. The detailed analysis of
the reflection-and-transmission problem for the Dirac electrons demonstrates essential
differences from the conventional Schr\"odinger case, due to the role of the Berry phase.
The reflection coefficient turns out to be nonanalytic function of the
transverse momentum $p_y$ vanishing as $|p_y|$ at $p_y \to 0$. 

We have presented a complete treatment of the chiral tunneling for both single and bilayer graphene in terms of a
classical phase space. This gives a natural explanation of complete transmission of normally
incident wave for single layer and its exponentially damped transmission in bilayer.
We have also demonstrated that, for the case of nonsymmetric n-p-n junction in single layer graphene,
there is total transmission for the normal incidence only, and other maxima are suppressed.
Our numerical studies show that for the case of bilayer there are always magic angles with
total transmission.

\section{Acknowledgments}

We are thankful to Sergey Dobrokhotov, Andrey Shafarevich, Anna Esina and Andrey Shytov for
helpful discussions.

We acknowledge financial support from the Stichting voor Fundamenteel Onderzoek der Materie (FOM),
which is financially supported by the Nederlandse Organisatie voor Wetenschappelijk Onderzoek (NWO).

\appendix

\section{\label{app::ad-ham}Effective Hamiltonians in adiabatic approximation}

In this part of the Appendix we show how to reduce in adiabatic approximation an initial matrix Hamiltonian
to a set of effective scalar Hamiltonians. Our consideration follows \cite{Berlyand87,Belov06}.

Let us consider an eigenvalue problem for an Hermitian matrix Hamiltonian $\hat{H}$,
\begin{equation}
 H(-ihd/dx,x)\Psi(x)=E\Psi(x) \; ,
\label{eq::stat-sch}
\end{equation}
where we assume that $\hat{p}_x=-ihd/dx$ acts first and $x$ acts second. In what follows we always assume this operator ordering. Let us introduce a vector operator $\hat{\chi}$ and a scalar wave function $\psi$ by the requirement
\begin{equation}
 \Psi(x)=\chi(-ihd/dx,x,h)\psi(x) \; .
\end{equation}
We want $\psi$ to satisfy an eigenvalue problem
\begin{equation}
 L(-ihd/dx,x,h) \psi(x)=E \psi(x) \; , \label{eq:appEffSchrodx}
\end{equation}
with an effective Hermitian Hamiltonian $\hat L$. Substituting (\ref{eq:appEffSchrodx}) in (\ref{eq::stat-sch}) we obtain:
\begin{equation}
 (\hat H\hat\chi-\hat\chi\hat L)\psi(x)=0.
\end{equation}
The last equality will be fulfilled for any $\psi(x)$ if the following operator equality holds:
\begin{equation}
\fl H(-ihd/dx,x)\chi(-ihd/dx,x,h)=\chi(-ihd/dx,x,h) L(-ihd/dx,x,h).
\end{equation}
We will solve it by passing to symbols of operators (see~\cite{Maslov72,Maslov81}):
\begin{equation}
  \textrm{smb}[A(-ihd/dx,x) B(-ihd/dx,x)] = A(p_x-ihd/dx,x) B(p_x,x) \; .
\end{equation}
Applying this formula to the above case, we obtain
\begin{equation}
 H(p_x-ihd/dx,x)\chi(p_x,x,h)=\chi(p_x-ihd/dx,x,h) L(p_x,x,h) \; .
\end{equation}
Let us expand this expression with respect to the parameter $h\ll 1$. In zeroth order we get
\begin{equation}
 H(p_x,x)\chi_0(p_x,x)=L_0(p_x,x)\chi_0(p_x,x) \; , \label{eq:appH0eigenvalue}
\end{equation}
where $L_0(p_x,x)=L(p_x,x,0)$, $\chi_0(p_x,x)=\chi(p_x,x,0)$. The first order term in the expansion gives
\begin{equation}
 -i\frac{\partial H}{\partial p_x}\frac{\partial \chi_0}{\partial x}+H\chi_1=
 -i\frac{\partial \chi_0}{\partial p_x}\frac{\partial L_0}{\partial x}+L_0\chi_1+\chi_0 L_1 \; ,
\end{equation}
where $L_1$ and $\chi_1$ are the first order terms in $L$ and $\chi$ with respect to $h$. The above expression can be rewritten as
\begin{equation}
 (H-L_0)\chi_1=i\frac{\partial H}{\partial p_x}\frac{\partial \chi_0}{\partial x}
-i\frac{\partial \chi_0}{\partial p_x}\frac{\partial L_0}{\partial x}+\chi_0 L_1 \; .
\end{equation}
Let us multiply the last equation by $\chi_0^\dagger$ from the left. Since both $H$ and $L_0$ are Hermitian matrices, we obtain
\begin{equation}
L_1=-i\chi_0^\dagger\frac{\partial H}{\partial p_x}\frac{\partial \chi_0}{\partial x}
+i\chi_0^\dagger\frac{\partial \chi_0}{\partial p_x}\frac{\partial L_0}{\partial x} \; ,
\label{eq:appL1solution}
\end{equation}
where we used the equality $\chi_0^\dagger \chi_0 = 1$.

\section{\label{app::sem-appr}Semiclassical approximation}

\subsection{$x$-representation}
To solve equation~(\ref{eq:appEffSchrodx}) we will use the semiclassical ansatz
\begin{equation}
 \psi(x)=e^{iS(x)/h}A(x,h) \; . \label{eq:appsolutionx}
\end{equation}
Then (\ref{eq:appEffSchrodx}) can then be rewritten as
\begin{equation}
 L(dS/dx-ihd/dx,x,h)A(x,h)=E A(x,h) \; . \label{eq:applongmomentumx}
\end{equation}
This equation can be expanded order by order in $x$ which to zeroth order gives the Hamilton-Jacobi equation
\begin{equation}
 L_0(dS/dx,x)=E \; . \label{eq:apphamiltonjacobix}
\end{equation}
From this equation we can determine the action $S(x)$, as is well known from classical mechanics~\cite{Goldstein02}. To first order in $h$ we obtain the equation
\begin{equation}
 -i\frac{\partial L_0}{\partial p_x}\frac{\partial A_0}{\partial x}+L_1 A_0-\frac{i}{2}\frac{\partial^2 L_0}{\partial p_x^2} \frac{d^2 S} {d x^2}A_0=0 \; , \label{eq:appexpfirstcorrectionx}
\end{equation}
where all terms should be evaluated at $p_x = dS/dx$. When multiplying by the amplitude $A_0$, equation~(\ref{eq:appexpfirstcorrectionx}) can be rewritten as
\begin{equation}
 -\frac{i}{2}\frac{d}{d x}\left(\frac{\partial L_0}{\partial p_x}A_0^2\right)+
\left(L_1+\frac{i}{2}\frac{\partial^2 L_0}{\partial p_x\partial x}\right)A_0^2=0 \; ,
\end{equation}
where the total derivative acts on both $x$ and $p_x = dS/dx$. This equation can be solved exactly to determine the amplitude
\begin{eqnarray}
 A_0&=&\left|\frac{\partial L_0}{\partial p_x}\right|^{-1/2}\exp\left[-i\int dx \left(\frac{\partial L_0}{\partial p_x}\right)^{-1}
\left(L_1+\frac{i}{2}\frac{\partial^2 L_0}{\partial p_x\partial x}\right) \right] \\
&\equiv&\left|\frac{\partial L_0}{\partial p_x}\right|^{-1/2}\exp\left[\int dx \left(\frac{\partial L_0}{\partial p_x}\right)^{-1}M\right] \; , \label{eq:appamplx}
\end{eqnarray}
where we have defined $M$. Using equation~(\ref{eq:appL1solution}) it can be written as
\begin{equation}
 M=-\chi_0^\dagger\frac{\partial H}{\partial p_x}\frac{\partial \chi_0}{\partial x}
+\chi_0^\dagger\frac{\partial \chi_0}{\partial p_x}\frac{\partial L_0}{\partial x}+\frac{1}{2}\frac{\partial^2 L_0}{\partial p_x\partial x} \; .
\label{eq:appexpMx}
\end{equation}

\subsection{$p$-representation}
We can also solve equation~(\ref{eq:appEffSchrodx}) by passing to $p$-representation~\cite{Maslov72,Maslov81}:
\begin{equation}
 L(\stackrel{1}{p_x},\stackrel{2}{ihd/dp_x},h)\widetilde\psi(p_x)=E\widetilde\psi(p_x) \; . \label{eq:appEffSchrodp}
\end{equation}
In this equation $p_x$ acts first, while $ih d/dp_x$ acts second, contrary to the case we considered before. To solve this equation, we use the semiclassical ansatz
\begin{equation}
 \widetilde \psi(p_x)=e^{-i\widetilde S(p_x)/h}\widetilde A(p_x,h) \; . \label{eq:appsolutionp}
\end{equation}
Similarly to equation~(\ref{eq:applongmomentumx}) equation~(\ref{eq:appEffSchrodp}) can be rewritten as
\begin{equation}
 L(p_x,d\widetilde S/dp_x+ihd/dp_x,h)\widetilde A(p_x,h)=E \widetilde A(p_x,h) \; ,
\end{equation}
When this equation is expanded order to order in $h$, one obtains to zeroth order the Hamilton-Jacobi equation
\begin{equation}
 L_0(p_x,d\widetilde S/dp_x)=E \; , \label{eq:apphamiltonjacobip}
\end{equation}
from which the action $\widetilde S(p_x)$ can be determined. The first order term becomes
\begin{equation}
 i \frac{\partial L_0}{\partial x} \frac{\partial \widetilde A_0}{\partial p_x} + L_1 \widetilde A_0 +
\frac{i}{2} \frac{\partial^2 L_0}{\partial x^2} \frac{d^2 \widetilde S}{d p_x^2} \widetilde A_0 +
i \frac{\partial^2 L_0}{\partial x \partial p_x} \widetilde A_0 = 0 \; ,
\end{equation}
where all terms have to be evaluated at $x = d\widetilde S/d p_x$. After multiplication by $\widetilde A_0$ one finds
\begin{equation}
 \frac{i}{2}\frac{d}{d p_x}\left(\frac{\partial L_0}{\partial x}\widetilde A_0^2\right)+
\left(L_1+\frac{i}{2}\frac{\partial^2 L_0}{\partial p_x\partial x}\right)\widetilde A_0^2=0 \; ,
\end{equation}
which can be solved exactly to give
\begin{eqnarray}
\widetilde A_0&=&\left|\frac{\partial L_0}{\partial x}\right|^{-1/2}\exp\left[i\int dp_x \left(\frac{\partial L_0}{\partial x}\right)^{-1}
\left(L_1+\frac{i}{2}\frac{\partial^2 L_0}{\partial p_x\partial x}\right) \right]\\
&\equiv&\left|\frac{\partial L_0}{\partial x}\right|^{-1/2}\exp\left[-\int dp_x \left(\frac{\partial L_0}{\partial x}\right)^{-1}M\right] \; . \label{eq:appamplp}
\end{eqnarray}
Using equation~(\ref{eq:appL1solution}) the quantity $M$ can then be written as
\begin{equation}
 M=-\chi_0^\dagger\frac{\partial H}{\partial p_x}\frac{\partial \chi_0}{\partial x}
+\chi_0^\dagger\frac{\partial \chi_0}{\partial p_x}\frac{\partial L_0}{\partial x}+\frac{1}{2}\frac{\partial^2 L_0}{\partial p_x\partial x} \; .
\label{eq:appexpMp}
\end{equation}

\subsection{Matching}
Since the solutions~(\ref{eq:appsolutionx}) and~(\ref{eq:appsolutionp}) come from the same equation, they should be related. To find out what this relation is, we look at the Fourier representation of~(\ref{eq:appsolutionx}), which is defined by
\begin{eqnarray}
% \psi(x)=\frac{1}{\sqrt{2\pi h}}\int_{-\infty}^\infty e^{ip_x x/h}\psi(p_x)dp_x\simeq e^{iS(x)/h}A_0(x) \; ,\\
 \widetilde \phi(p_x)=\frac{1}{\sqrt{2\pi h}}\int_{-\infty}^\infty e^{i(S(x)-p_x x)/h} A_0(x)dx
\end{eqnarray}
Since the parameter $h$ is assumed to be small, we can calculate this integral using the stationary phase method. The result is
\begin{equation}
 \widetilde\phi(p_x)=\frac{A_0(x_s)}{\sqrt{|S''(x_s)|}}e^{i(S(x_s)-p_x x_s)/h+i\,\mathrm{sgn}(S''(x_s))\,\pi/4}, \label{eq:appFTsolx}
\end{equation}
where the point $x_s = x_s(p_x)$ at which the phase is stationary is to be found from the equality
\begin{equation}
  S'(x_s)=p_x \; . \label{eq:appcondstatphase}
\end{equation}
From the comparison of (\ref{eq:appsolutionp}) and (\ref{eq:appFTsolx}) we find \cite{Maslov72,Maslov81}
\begin{equation}
  \widetilde\phi(p_x) = e^{i\,\mathrm{sgn}(S''(x_s))\,\pi/4}\widetilde\psi(p_x) \; . \label{eq:apprelationpxrep}
\end{equation}

\section{\label{app::bohr-somm}Bohr-Sommerfeld quantization rule}

\begin{figure}
\begin{center}
\includegraphics[width=10cm]{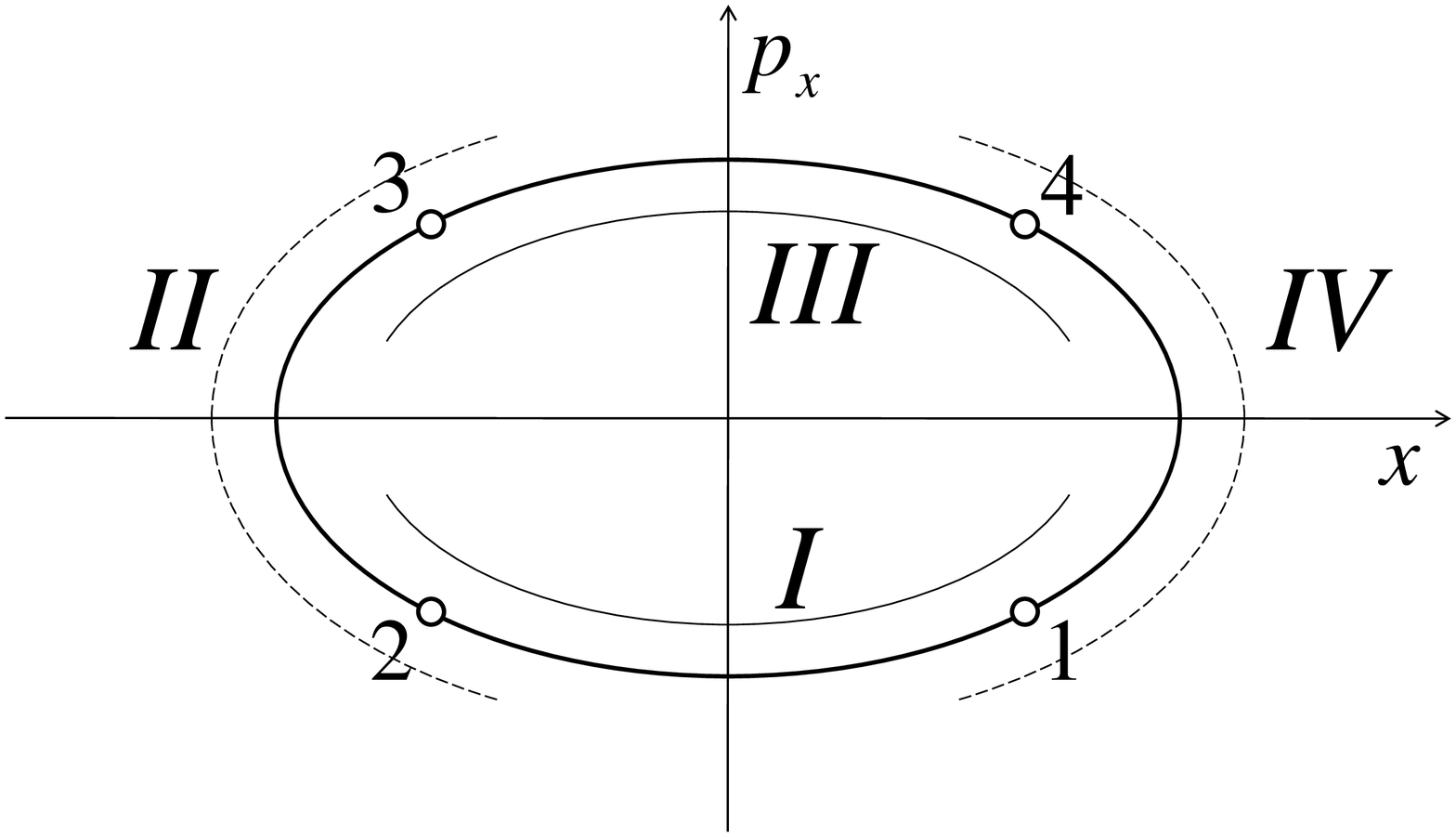}
\end{center}
\caption{The phase space of a classical particle is covered by maps $I$, $II$, $III$, $IV$. Regular maps $I$ and $III$ can be uniquely projected onto the $x$-axis, while singular maps $II$ and $IV$ can be uniquely projected onto the $p_x$-axis. The maps are chosen to overlap. }
\label{fig:WKBMatlas}
\end{figure}

Now let us consider the phase space which is shown in figure~\ref{fig:WKBMatlas}.
The part $I$ of the phase trajectory can be projected onto the $x$-axis. Therefore we can use a standard WKB ansatz in this region. On the contrary, in region $II$ we can use a standard WKB ansatz in $p$-representation. Since regions $I$ and $II$ overlap, functions are related according to (\ref{eq:apprelationpxrep}). Since  $S''(x_2)<0$, we have $\psi_{I}(x)\to \widetilde\psi_{II}(p_x) e^{-i\pi/4} $.

It is easily seen that the above reasoning can also be applied to regions $II$ and $III$. Since $S''(x_3)>0$ we obtain $\psi_{III}(x)\to\widetilde\psi_{II}(p_{x}) e^{i\pi/4}$, or
\begin{equation}
  \psi_{I}(x) \to \psi_{III}(x) e^{-i \pi/2} \; .
\end{equation}
So, passing the turning point lying in the region $II$ we have picked up an extra factor $\exp(-i \pi/2)$. Since $S''(x_4)<0$ and $S''(x_1)>0$ we pick up another factor of $\exp(-i \pi/2)$ when we go through region $IV$ and pass the second turning point,
\begin{equation}
  \psi_{III}(x) \to \psi_{I}(x) e^{-i \pi/2} \; .
\end{equation}
In passing one full turn along the circle, one therefore sees that $\psi_{I}(x) \to \psi_{I}(x) e^{-i \pi}$. The wavefunction should be single-valued, which means that the exponent should be a multiple of $2 \pi$. From equations~(\ref{eq:appsolutionx}) and~(\ref{eq:appamplx}) we therefore find
\begin{equation}
  \frac{1}{h}\oint p_x dx-\phi_B-\pi=2\pi n \; ,  \label{eq:appPrequant}
\end{equation}
where $p_x = S'(x)$ is to be found from the Hamilton-Jacobi equation~(\ref{eq:apphamiltonjacobix}). The quantity $\phi_B$ is the Berry phase, defined by
\begin{equation}
  \phi_B=i\oint dx \left(\frac{\partial L_0}{\partial p}\right)^{-1}
  \left(-\chi_0^\dagger\frac{\partial H}{\partial p_x}\frac{\partial \chi_0}{\partial x}
  +\chi_0^\dagger\frac{\partial \chi_0}{\partial p_x}\frac{\partial L_0}{\partial x}+\frac{1}{2}\frac{\partial^2 L_0}{\partial p\partial x}\right) \; .
\end{equation}
Equation~(\ref{eq:appPrequant}) can be rewritten as the Bohr-Sommerfeld quantization rule
\begin{equation}
  \frac{1}{2\pi}\oint p_x dx=h\left(n+\frac{1}{2}+\frac{\phi_B}{2\pi}\right) \; , \label{eq:bohrsommerfeld}
\end{equation}
Looking back at the above derivation one sees that the term $1/2$ can be written as $\nu/4$, where $\nu$ is the number of turning points (Maslov index in this particular case)~\cite{Maslov81}.

\section{\label{app::parcas}Asymptotic expansions of parabolic cylinder functions in different Stokes sectors}

For completeness we placed in this section the asymptotic expansions of the parabolic cylinder functions in different Stokes sectors at $|z|\to\infty$ according to \cite{WolframFunctions}:
\begin{equation}
\fl
 D_\nu(z)\sim \left\{\begin{array}{ll}
e^{-z^2/4}z^\nu(1+O[z^{-2}]), & -\pi/2<\textrm{arg}\,(z)\leq\pi/2,\\
e^{-z^2/4}z^\nu(1+O[z^{-2}])-\frac{e^{z^2/4-i\pi\nu}\sqrt{2\pi}z^{-\nu-1}}{\Gamma(-\nu)}(1+O[z^{-2}]), & \textrm{arg}\,(z)\leq-\pi/2\\
e^{-z^2/4}z^\nu(1+O[z^{-2}])-\frac{e^{z^2/4+i\pi\nu}\sqrt{2\pi}z^{-\nu-1}}{\Gamma(-\nu)}(1+O[z^{-2}]), & \textrm{arg}\,(z)>\pi/2
\end{array}\right.
\end{equation}
We assume that $-\pi<\textrm{arg}\,(z)<\pi$.

%\bibliographystyle{unsrt}
%\bibliography{thispaper}

\end{document}